\documentclass[]{aa}  
\usepackage{ae,aecompl}
\usepackage{adjustbox}
\usepackage{lineno}
\usepackage{graphicx}
\usepackage{txfonts}
\usepackage{hyperref}
\usepackage{xcolor}

\hypersetup{colorlinks,linkcolor={blue},citecolor={blue},urlcolor={blue}}  
\usepackage{graphicx}
\usepackage{epsf,palatino,url,wrapfig,array,setspace,amsmath,amssymb,fancyhdr,multirow,lscape,appendix,rotating}
\usepackage{graphics,epsfig,txfonts}
\usepackage{threeparttable}
 
\begin{document}

    \title{Electromagnetic signatures of gradual magnetic dissipation in jets from supermassive black holes}
   \titlerunning{Emission from striped blazar jets}
   \author{S. Dimitrakoudis \inst{1}, M. Petropoulou  \inst{1,2}, S. I. Stathopoulos \inst{3} and D. Giannios \inst{4}  
          }       
   \institute{Department of Physics, National and Kapodistrian University of Athens, University Campus Zografos, GR 15784, Athens, Greece, \email{sdimis@phys.uoa.gr}
 \and
    Institute of Accelerating Systems \& Applications, University Campus Zografos, Athens, Greece
 \and
    Deutsches Elektronen-Synchrotron DESY, Platanenallee 6, 15738 Zeuthen, Germany
 \and
    Department of Physics, Purdue University, 525 Northwestern Avenue, West Lafayette, IN 47907, USA}
    
   \date{Received ... ; accepted ...}

 
  \abstract
   {Blazar jets -- collimated and relativistic plasma outflows viewed at small angles -- are prolific sources of non-thermal radiation.  
   }
   {We develop a numerical framework for calculating the emission from an extended blazar jet where gradual energy dissipation, mediated through magnetic reconnection, takes place across a wide range of distances from the supermassive black hole.}
   {We adapt the open-source radiative transfer code {\tt LeHaMoC} that self-consistently calculates photon spectra from  relativistic charged particles that occupy a spherical blob
   to treat the emission from an extended jet. The jet is treated as a series of segments that interact with each other through radiative transfer. The effects of radiation spillage are calculated iteratively throughout the jet, accounting for relativistic effects arising from the relative motion between jet segments, until a steady state is reached.}
   {We find that this approach produces some distinct differences in the total spectral energy distribution compared to the results obtained by adding the emission escaping from each constituent segment of the jet (optical thin regime), increasing the part of the spectrum affected by inverse Compton scattering by a factor of three. However, the latter is always far less luminous than the synchrotron component, even when enhanced by external radiation fields. }
   {Gradual magnetic dissipation in jets from supermassive black holes leads to nearly universal, synchrotron-dominated photon spectra. The numerical framework developed here is readily adaptable to a broad range of jet dissipation scenarios and applications to extended jet emission.}

   \maketitle
   \nolinenumbers
%

\section{Introduction}
\label{sec:intro}
Relativistic jets from accreting black holes are ubiquitous sources of electromagnetic radiation. Their emission is powered by some energy dissipation mechanism that transforms part of the jet energy (stored in kinetic or electromagnetic form) into energy carried by a relativistic particle population. The latter step requires an in situ particle acceleration mechanism operating at large distances from the black hole~\citep[for a review, see][]{2020NewAR..8901543M}.

Contrary to Galactic stellar-scale jets (microquasars) where imaging can provide spatial information about the emission at different energies \citep[e.g.][]{Blundell2001, Hess_SS433, LHAASO_BH_2024}, in jets from active galactic nuclei (AGN) the prevailing method for pinpointing the location of the high-energy (HE, namely X-ray and $\gamma$-ray) time variable emission is the modeling of the spectral energy distribution (SED). In the case of blazar jets -- blazars are active galaxies with relativistic jets that are viewed at small angles ($\theta < 10^{\rm o}$) with respect to our line of sight -- the location of the HE emitting region, which is known  as 
the ``blazar zone'', is uncertain and depends on the underlying emission model (e.g.  $\sim 0.001$ to $\sim 30$ pc, \cite{Dermer1994, Sikora1994, Georganopoulos2003, Ghisellini2008, Agudo2011, Potter2012II}).

Blazar jets shine across the electromagnetic spectrum because of various dissipative processes that can transfer energy from the bulk of the jet to a fraction of plasma particles that end up becoming relativistic. One possible mechanism is magnetic reconnection driving continuous energy dissipation throughout the jet \citep[]{Spruit2001, 2002A&A...387..714D, Giannios2009}. In this scenario, the jet is formed by open magnetic field lines whose polarity reverses over time scales comparable to the orbital time in the inner parts of the disk~\citep[e.g.,][]{2015MNRAS.446L..61P, 2018ApJ...864...52K, 2019MNRAS.490.4811C}. Such magnetic field reversals, in a disk with differential rotation that is the driving agent of a striped wind jet, have been seen in general relativity magnetohydrodynamic simulations of accretion disks with turbulence driven by magneto-rotational instabilities \citep[e.g.,][]{Kaufman2023, Das2024}. This results in the jet being `striped', i.e., having an alternating toroidal field polarity with current sheets forming between the stripes of different magnetic polarity. Magnetic reconnection takes place in those current sheets, powering acceleration (of the bulk flow and of individual particles) in the jet \citep{2017MNRAS.468.3202B, Giannios2019}.

The striped jet model has been discussed in the context of Gamma-Ray Bursts (GRB) \citep{2017MNRAS.468.3202B} and more recently of blazars. More specifically, \cite{2021MNRAS.502.1145Z} studied the effects of bulk acceleration beyond parsec scales, focusing on the low-energy jet emission (namely, the radio spectral index and core-shift effect\footnote{The frequency-dependent apparent position of the radio core.}). An application of this framework to the multi-messenger flare of blazar TXS~0506+056 was presented by \cite{2025MNRAS.537.3895D}. Their work focuses on the multi-wavelength emission arising from a blob located beyond the transition zone from a magnetic-dominated to a kinetic-dominated jet. 

In this paper we study for the first time the broadband electromagnetic signatures arising from non-thermal electrons accelerated in blazar jets across a wide range of distances under the continuous magnetic dissipation scenario~\citep[]{Spruit2001, Giannios2009}. Our goal is to explore the properties of the blazar jet continuous emission and the impact of physical parameters, such as the initial magnetization of the jet, the black hole mass, and jet power. For the jet dynamics, we follow the semi-analytical model laid out by \cite{Giannios2019}. Building upon this, we construct a numerical scheme to compute the jet emission over a broad range of distances, incorporating gradual energy dissipation and bulk acceleration. A key novelty of our approach is the inclusion of inter-segment photon exchange, whereby a fraction of photons escapes from one segment and interacts with adjacent regions. Additionally, we account for the relativistic boosting of radiation resulting from the differential motion between neighboring jet segments. 

This paper is structured as follows. We outline the continuous magnetic dissipation jet model in Sect.~\ref{sec:Jet_model}. We present the numerical approach to discretize the jet in Sect.~\ref{sec:numerical-model}. In Sect.~\ref{sec:results} we present our results and conclude with a discussion in Sect.~\ref{sec:conclusions}.
 
\section{A dissipative jet model}
\label{sec:Jet_model}
The jet is described by two conserved quantities, namely the mass outflow rate $\dot{M}$ and the total (kinetic and electromagnetic) power $L =L_{\rm kin} + L_{\rm Poyn} = \Gamma(z) \dot{M} c^2 (1+\sigma(z))$, where $\sigma$ is the magnetization of the jet and is much larger than unity for Poynting flux-dominated jets. Given that $\dot{M}$ and $L$ are independent of $z$, we can write $\Gamma(z)  (1+\sigma(z)) = \Gamma_{\infty}$ with $\sigma_{\infty} \ll 1$, where the subscript $\infty$ is used to indicate asymptotic values of the variables. Therefore, we may infer the outflow rate of the rest-mass energy as 
\begin{equation}
    \dot{M}= \frac{L}{\Gamma_\infty c^2},
    \label{eq:Mdot_out}
\end{equation}
if the terminal Lorentz factor of the jet is known. 

\cite{Giannios2019} computed the bulk acceleration of the jet flow due to magnetic dissipation at reconnection current sheets, ${\rm d} u/ {\rm d} z$, where $u = \Gamma \beta$ is the jet's (dimensionless) spatial component of the four-velocity. Its gradient is related to the properties of the width $l$ distribution of stripes in the jet (see, e.g., Eqs. 16 and 17 in \cite{Giannios2019}). For a power-law distribution $P(l) \propto l^{-a}, l \ge l_{\min}$, the four-velocity normalized to its terminal value, $\chi = u(z)/u_{\infty} \approx  \Gamma(z)/\Gamma_{\infty} $, is the root of the following equation,
\begin{eqnarray}
\label{eq:chi-zeta-C}
\frac{(1-\chi)^{3-k}} {k-3} - \frac{2(1-\chi)^{2-k}}{k-2} + \frac{(1-\chi)^{1-k}}{k-1} = \zeta + C, 
\end{eqnarray}
where $C = (k-3)^{-1} - 2 (k-2)^{-1} + (k-1)^{-1}$,  $k = (1-3a)/(2-2a)$,  $\zeta = 2z/z_{\rm rec}$, $z_{\rm rec} = u_{\infty}^2 l_{\min}/\beta_{\rm rec} = 10\, \Gamma^2_{\infty} l_{\min}/\beta_{\rm rec, -1}$, and $\beta_{\rm rec}$ is the reconnection rate. For blazars, $z_{\rm rec} \sim 10^6~R_{\rm g} \, \Gamma^2_{\infty, 1} l_{\min, 3}$, where $l_{\min, 3} = l_{\min}/(10^3 R_{\rm g})$, $\Gamma_{\infty, 1} = \Gamma_{\infty}/10$, and $R_{\rm g} = GM/c^2$. The solution $\chi(\zeta)$ can be used to determine the rate of power dissipation and magnetization. 

\begin{figure}
\centering
\includegraphics[width=0.45\textwidth]
{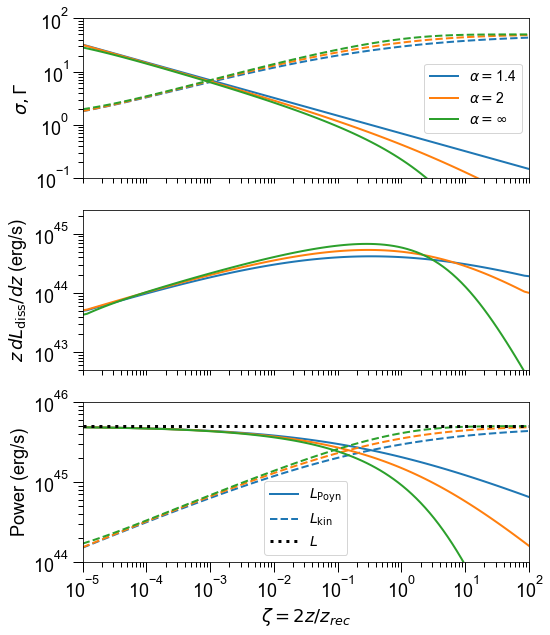}
\caption{Jet parameters as a function of distance (normalized to $z_{\rm rec}/2$) for three values of $\alpha$ indicated by different colors. Top panel: magnetization (solid lines) and bulk Lorentz factor (dashed lines). Middle panel: dissipated power per logarithmic interval $dz$. Bottom panel: kinetic and Poynting jet power. 
A jet with total power $L=5 \cdot 10^{45}$~erg s$^{-1}$ and $\Gamma_{\infty}=50$ was assumed.}
\label{fig:sigma_dLdz}
\end{figure}

The differential energy dissipation rate in the jet between the distances $z$ and $z+{\rm d}z$ is written as,
\begin{equation} 
\frac{{\rm d}L_{\rm diss}} {{\rm d}z} = \frac{{\rm d}\Gamma} {{\rm d} z} \dot{M} c^2 = \frac{L}{\Gamma_{\infty}} \frac{{\rm d}\Gamma} {{\rm d} z} = L \frac{{\rm d}\chi} {{\rm d} z}. 
\label{eq:dLdz}
\end{equation} 
The magnetization at distance $\zeta$ is given by
\begin{equation} 
\sigma(\zeta) = \frac{\Gamma_{\infty}}{\Gamma} - 1 = \frac{1}{\chi(\zeta)} -1. 
\label{eq:sigma}
\end{equation}
In the special case of $a=1.4$ the solution to Eq.~(\ref{eq:chi-zeta-C}) is explicit and reads,
\begin{equation}
\chi(\zeta) = {3\zeta - (3\zeta)^{2/3} +(3\zeta)^{1/3} \over 1+3\zeta}.
\end{equation}
For all other values of $a$, $L_{\rm diss}(z)$ and $\sigma(z)$ are computed numerically. Figure~\ref{fig:sigma_dLdz} shows the evolution of $\sigma(z)$, $\Gamma(z)$, and $z {\rm d}L_{\rm diss}/{\rm d}z$ for three values of $a$. For a wide range of distances, we find that $\sigma \propto z^{-1/3}$, $\Gamma \propto z^{1/3}$, and $z {\rm d}L_{\rm diss}/{\rm d}z \propto z^{1/3}$. 

Knowing the magnetization $\sigma$ of the jet at distance $z$ we can compute the Poynting luminosity 
\begin{equation}
L_{\rm Poyn} = \sigma L_{\rm kin} =\frac{\sigma}{1+\sigma} L,
\label{eq:Lpoyn}
\end{equation}
which is plotted as a function of distance $z$ in the lower panel of Figure~\ref{fig:sigma_dLdz} together with $L_{\rm kin}$ for three values of $\alpha$.  From the jet Poynting luminosity (see Eq.~A1 in \cite{stathopoulos2026}) we derive the (comoving)\footnote{Primed quantities are measured in the fluid rest frame. Because the bulk velocity of the jet changes with $z$ there is no unique rest frame for the whole jet.} magnetic field strength at that distance,
\begin{equation}
B'(z) =
\frac{1}{R \Gamma }\sqrt{\frac{8 L\, \sigma}{\beta c(1+\sigma)}},
\label{eq:B}
\end{equation}
where $R(z) = \Delta z/2$ is the cross-sectional radius of a conical jet at that distance (see Eq.~\ref{eq:dz}), which is Lorentz invariant. As long as $\sigma > 1$, the comoving magnetic field strength is independent of the magnetization and scales as $B' \propto z^{-4/3}$. The comoving number density of the jet plasma at distance $z$ can easily be derived through the mass continuity equation,
\begin{equation}
n'_{\rm bulk}(z) = \frac{\dot{M}}{\bar{m} \pi R^{2} \Gamma \beta c } \propto z^{-7/3}
\label{eq:nbulk}
\end{equation}
where $\bar{m} = m_{\rm p}/2$ assuming a electron-proton plasma.

A fraction $f_{\rm rec}$ of the dissipated energy at distance $z$ is transferred to a population of non-thermal electrons accelerated by magnetic reconnection. Particle-in-cell simulations (PIC) of magnetic reconnection in the relativistic regime ($\sigma \gtrsim 1$) have provided a detailed view of the non-thermal particles accelerated in reconnection layers and their properties. For example, PIC simulations in two dimensions (2D) have shown that both the fraction of energy transferred to non-thermal particles $f_{\rm rec}$, and the power-law index of the particle distribution $s$, depend on $\sigma$ \citep[e.g.][]{Guo2014, Sironi2014, Sironi2015}. Here, we use the empirical relation found by \citet{Sironi2015} (for zero guide field) which reads 
\begin{equation}
    f_{\rm rec} \simeq \xi \frac{\sigma}{\sigma+2}
    \label{eq:frec}
\end{equation}
where $\xi \sim 0.27$ for electron-proton plasmas and $\xi \sim 0.55$ for electron-positron plasmas. In this work we choose $\xi = 0.27$ assuming a jet composed of electrons and protons, although we are only considering the radiative signatures of non-thermal leptons in this work.

Particle acceleration in reconnection layers is a fast process. Here, we consider it instantaneous compared to all other timescales, and use an injection term $Q_{\rm e}(\gamma',z')$ to describe the injection rate of non-thermal electrons in a jet segment centered at $z'$ with height $\Delta z'$,
\begin{equation} 
Q'_{\rm e}(\gamma', z') = Q'_0(z') \, \gamma^{' -s}, \, \gamma'_{\min} \le \gamma' \le \gamma'_{\max},
\label{eq:Qe}
\end{equation} 
where $\gamma'_{\min}$,  $\gamma'_{\max}$ are the minimum and maximum Lorentz factors of the distribution, respectively, $s$ is the power-law slope, and $Q'_0$ is a normalization parameter. For simplicity, we assume that $\gamma'_{\min} = 1$, and note that our results are not strongly impacted by this choice.
The maximum Lorentz factor is evaluated as the minimum of two values: the size-limited Lorentz factor 
\begin{equation}
\gamma_{\rm H} = \frac{eB'R_{\rm avg}}{\Gamma m_{\rm e} c^2},
\label{eq:gamma-Hillas}
\end{equation}
where $R_{\rm avg}$ is the average distance from the center of the segment to its sides in its own rest frame (see Appendix \ref{app:segments}),
and the synchrotron-limited Lorentz factor (where the synchrotron loss timescale, $t_{\rm syn} = 6 \pi m_{\rm e}c/(\sigma_{\rm T} B^{'2}\gamma')$, equals the acceleration timescale, $t_{\rm acc} = 
\eta_{\rm acc} m_{\rm e} \gamma' c^2/(eB'c)$~\citep{2021MNRAS.501.4092R, 2023MNRAS.518.2719P}. PIC simulations find $\eta_{\rm acc} \sim 10$; however, past studies have shown that to recover blazar-like spectra one would have to assume longer acceleration timescales. Motivated by those studies, here we adopt $\eta_{\rm acc}=10^4$. The synchrotron-limited Lorentz factor is
\begin{equation}
\gamma'_{\rm rad} = \sqrt{6 \pi e  \over \eta_{\rm acc} \sigma_T B'}.
\label{eq:gamma-rad}
\end{equation}
For all parameters considered in this work, we find $\gamma'_{\max} = \gamma'_{\rm rad}\propto z^{2/3}$. 
The average Lorentz factor of the particle distribution can then be derived,
\begin{equation}
\langle \gamma' \rangle = \frac{\int {\rm d}\gamma' \gamma^{'-s+1}}{\int {\rm d}\gamma'  \gamma^{' -s}},
\end{equation}
where the power-law distribution at injection was used as a weighting factor. 

We implement the $\sigma$-dependence of the power-law index $s$ as per \citet{Werner2016} for the relativistic regime ($\sigma \gtrsim 1)$. For lower values of $\sigma$ we have used the results of \citet{Ball2018}, which seamlessly transitions to the values of \citet{Werner2016} for intermediate values of $\sigma$ when the plasma parameter $\beta$ (defined as the ratio of proton thermal pressure to magnetic pressure) is low, as for a cold jet plasma (see also Fig.~1 in \cite{stathopoulos2026}). 
The discrete values for the index $s$ obtained from the literature are then fitted with a polynomial function, which is used in our numerical jet model:
\begin{eqnarray}
s = 0.0595 \, x^6 - 0.4342\, x^5 + 0.9628\, x^4 - 0.3155\, x^3 \nonumber \\ 
- 0.8336\, x^2 - 0.3883 \, x + 2.5849
\label{eq:s_fit}
\end{eqnarray}
where $x = \mathrm{log_{10}}(\sigma)$. If $\sigma < 0.1$ or $s > 4$, indicative of inefficient acceleration to high energies, we fix $s=4$.

The power injected into non-thermal electrons in a jet segment at distance $z$ with height $\Delta z$ is written as 
\begin{equation}
L_{\rm diss, e}(z) = \int^{z+\Delta z}_{z } f_{\rm rec}(z) \frac{{\rm d}L_{\rm diss}} {{\rm d}z} {\rm d}z = L\int^{z+\Delta z}_{z}  f_{\rm rec} (z) \frac{{\rm d}\chi} {{\rm d}z} {\rm d}z. 
\label{eq:Le}
\end{equation}
This is related to the (isotropic) injection power measured in the segment's comoving frame as 
\begin{equation} 
L'_{\rm e} \approx \frac{L_{\rm diss, e}}{3 \Gamma^2}
\label{eq:L_inj_nt}
\end{equation}
where a beaming half-angle of $1/\Gamma$ was used \citep{2021MNRAS.501.4092R}; for more details see Appendix~A in \cite{stathopoulos2026}. Moreover, the injection power $L'_{\rm e}$ is related to the injection rate $Q'_{\rm e}(\gamma', z')$ as 
\begin{equation}
L'_{\rm e} = m_{\rm e} c^2 \int^{z'+\Delta z'}_{z'} Q'_0(z') \,  {\rm d}z' \int_{\gamma'_{\min}}^{\gamma'_{\max}(z')} \gamma^{-s(\sigma(z')) + 1} {\rm d} \gamma',
\end{equation}
which can be solved for $Q'_0(z')$ if the relation between $s$ and $z'$ is known. 

\begin{table}
\caption{List of model parameters. }
\label{tab:params}
\begin{center}
\begin{tabular}{l c c}
\hline
 Parameter & \multicolumn{2}{c}{Values} \\
           & Baseline & Alternate \\ 
\hline
Input &  &\\ 
\hline 
$M$ ($M_\odot$)$^a$ & $10^8$ & $10^7$ \\
$z_0$ ($R_{\rm g}$)$^b$ & 100 & - \\
$l_{\min}$ ($R_{\rm g}$)$^c$ & 400 & 40 \\
$a$ $^d$ & 1.4  & - \\
$\theta_{\rm j}$ $^e$ & $3^{\circ}$  & $10^{\circ}$  \\
$\theta_{\rm obs}$ $^f$ & $0^{\circ}$ & $15^{\circ}$  \\
$\Gamma_\infty$ $^g$ & 50 & 10\\
$L$ (erg/s) $^h$& $5 \cdot 10^{45}$ & $5 \cdot 10^{43}$, $5 \cdot 10^{47}$\\
$\beta_{\rm rec}$ $^i$ & 0.1 & - \\ 
$\eta_{\rm acc}$ $^j$ & $10^4$ & $10^2$ \\ 
\hline
Derived \\ 
\hline
$z_{\rm rec}$ $^k$ & 48~pc  \\
$\dot{M}$ $^l$ & Eq.~\ref{eq:Mdot_out} \\
$B'$ $^m$ & Eq.~\ref{eq:B} \\
$\gamma'_{\max}$ $^n$ & Eq.~\ref{eq:gamma-rad} \\
$s$ $^o$ & Eq.~\ref{eq:s_fit} \\
$L'_{\rm e}$ $^p$ & Eq.~\ref{eq:Le} \\
$z$ $^q$ & $5\cdot10^{-4}$pc  - $10^{3}$pc \\
\hline
\end{tabular}
    \begin{tablenotes}
      \small
      \item
      Symbol description: $^a$ Black hole mass, $^b$ distance from the center of the black hole to the start of the first segment of the jet, $^c$  minimum stripe width, $^d$ power-law index of the width distribution of stripes, $^e$ jet's opening angle as seen in the observer's frame, $^f$  observation angle to the jet's axis, $^g$ terminal jet bulk Lorentz factor, $^h$ total power in the jet, $^i$ reconnection rate, $^j$ acceleration efficiency, $^k$ a distance along the jet determined by the reconnection time of its stripes, $^l$ mass outflow rate, $^m$ magnetic field in the comoving frame, $^n$  maximum electron energy in the comoving frame, $^o$  slope of the accelerated electron distribution, $^p$ total power deposited in accelerated electrons in the comoving frame, $^q$ distance along the jet.
    \end{tablenotes}
\end{center}
\end{table}

\begin{figure*}
\centering
\includegraphics[width=0.9\textwidth]
{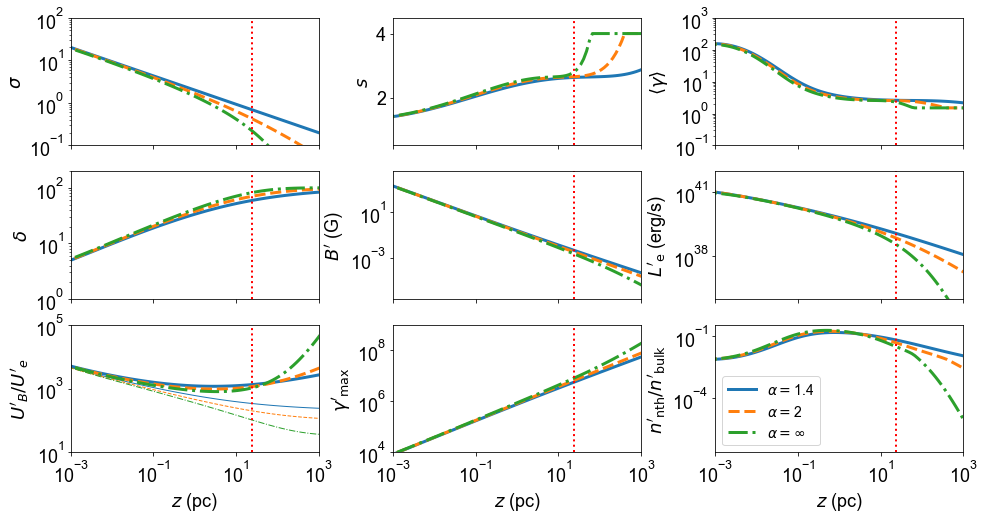}
\caption{Parameters of the model as a function of distance $z$ along the jet for three values of $\alpha$ (slope of the power-law distribution of width stripes). Vertical dotted lines indicate the distance $z_{\rm rec}/2$ for stripes with minimum width $l_{\min} = 400 \, R_{\rm g}$ for $M = 10^8 M_\odot$. A jet with total power $L = 5 \cdot 10^{45}$~erg~s$^{-1}$ and $\Gamma_{\infty}=50$ was assumed. Thin lines in the lower left panel show results for constant $f_{\rm rec}=0.27$. Primed quantities are measured in the respective segment's rest frame.
}
\label{fig:model-params}
\end{figure*}

In Figure \ref{fig:model-params} we present the dependence of several physical quantities defined in this section on the distance along the jet for three values of the power-law slope of the distribution of width stripes. For this example, we considered a baseline jet model with total power $L=5 \cdot 10^{45}$~erg s$^{-1}$ and $\Gamma_{\infty}=50$ (for the complete parameter list, see Table~\ref{tab:params}). The evolution of $\sigma$ and $\Gamma$ of the bulk flow with distance drives the dependence of all other parameters, such as $B', \gamma'_{\max}, s, L'_{\rm e}$, which are needed for calculating the non-thermal radiation emitted by each fluid element. The non-thermal electron number density (lower right panel) is at most 10\% of the jet plasma number density. Additionally, the energy density of the non-thermal electron population is $\lesssim 10^{-3} U'_B$, suggesting synchrotron-dominated emission arising from each jet segment (lower left panel). At distances beyond $z_{\rm rec}$ the ratio $U'_{\rm B}/U'_{\rm e}$ increases again as a result of the decreasing fraction $f_{\rm rec}$. Indeed, if $f_{\rm rec}=0.27$ were taken to be constant, as typically assumed in similar studies \citep{2017MNRAS.468.3202B, 2021MNRAS.502.1145Z, 2025MNRAS.537.3895D}, the energy density ratio would decrease asymptotically reaching a constant value at $z\gg z_{\rm rec}$ (see thin lines in the lower left panel of Fig.~\ref{fig:model-params}).

\section{Description of the numerical jet model}
\label{sec:numerical-model}

\begin{figure*}
\resizebox{\hsize}{!}{\includegraphics[clip=true]{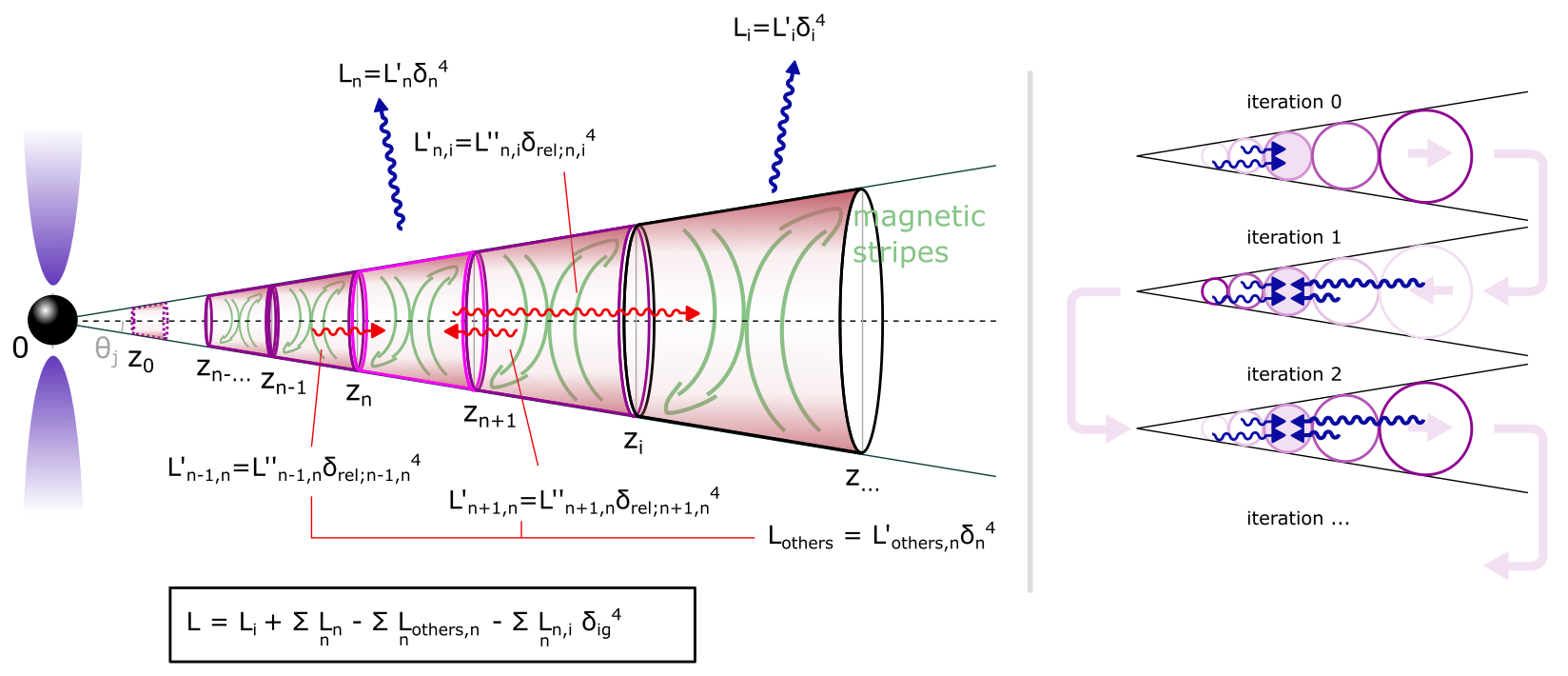}}
\caption{Schematics of the adopted numerical framework. Left panel: Sketch of a jet (not in scale) modeled as a series of segments inscribed within a cone, with an initial pseudo-parabolic bulk flow acceleration region represented by the border of the first segment (shown at $z_0$). The black hole, fed by an accretion disk, is located at $z=0$. The total luminosity we calculate at each point in the jet needs to account for the radiation from and between previous segments. Right panel: Schematic of our calculations from segment to segment, in repeated iterations going from the upstream end of the jet to the downstream end and back, until convergence is reached.
}
\label{fig:sketch}
\end{figure*}

The continuous jet is discretized into segments, each of which has its own rest frame, since $\Gamma$ increases with $z$. $\Delta z$ is directly proportional to $z$, so that each segment inscribes a sphere in the observer's frame (Appendix \ref{app:segments}). Each segment can be considered homogeneous, carrying part of the total jet power, which translates to a magnetic field $B'$ and a population of relativistic electrons injected by the acceleration process with a total luminosity $L'_{\rm e}$. 
The interaction of these elements can be modeled by solving their respective coupled kinetic equations within each segment, accounting for synchrotron radiation, inverse Compton scattering, synchrotron self-absorption, and $\gamma\gamma$ pair production. In all calculations, we assume that there is no transport of charged particles between segments or escape to the external medium.  On the contrary, photons emitted by one segment serve as an external source of radiation for the other segments of the jet. 

The steps for calculating the jet emission are shown in Figure~\ref{fig:sketch}. Each segment (inscribing a sphere but actually a frustum) receives a background photon density from each other sphere, deboosted by their relative Doppler factor; in every segment's frame, all the other segments, either in front of it or behind it, appear to be moving away from it. This is used within the model to calculate all interactions and ultimately the SED emerging from a given segment. This is then transformed to the observer's frame, boosted by the Doppler factor of that segment. When adding up the SEDs of the previous segments (boosted by their respective Doppler factors), their contributions to the background photon densities of other segments are subtracted, to avoid double-counting them. The model goes through a number of iterations, calculating the SED of each segment from one side of the jet to the other and then back again, until convergence is reached. That usually happens quickly enough, by the fourth iteration (Appendix \ref{app:iterations}).      

The numerical calculations are performed with the open-source software {\tt LeHaMoC}\footnote{\url{https://github.com/mariapetro/LeHaMoC/}} \citep{Stathopoulos2023}, adapted to calculate the pancake-like segments in an iterative fashion. Once the steady state SED is calculated for the first segment we must consider its effect on the next one. Since we are using {\tt LeHaMoC} in a purely leptonic model, the kinetic equations for the $i$-th segment\footnote{All variables are defined in the $i$-th segment's rest frame. We removed the primes to simplify the notation.} will be, for electrons and positrons:
 \begin{equation}
 {\partial N_{\rm e,i} \over \partial t} + {N_{\rm e,i} \over t_{\rm e,esc,i}} + \mathcal{L}_{\rm e,i}^{\rm syn} + \mathcal{L}_{\rm e,i}^{\rm ics} = \mathcal{Q}_{\rm e,i}^{\rm inj} + \mathcal{Q}_{\rm e,i}^{\gamma\gamma},  
 \end{equation}
and for photons:
 \begin{equation}
 {\partial N_{\rm \gamma,i} \over \partial t} + {N_{\rm \gamma,i} \over t_{\rm \gamma,esc,i}} + \mathcal{L}_{\gamma,i}^{\rm ssa} + \mathcal{L}_{\gamma,i}^{\gamma\gamma} = \mathcal{Q}_{\rm \gamma,i}^{\rm syn} + \mathcal{Q}_{\rm \gamma,i}^{\rm ics} + \sum_n \mathcal{Q}_{\rm \gamma,n,i}^{\rm xseg},  
 \end{equation}
where $N_{\rm e} \equiv {\rm d}N_{\rm e}/{\rm d}\gamma$, $N_{\gamma} \equiv {\rm d}N_{\gamma}/{\rm d}\nu$, $\mathcal{L}$ are particle/photon loss rates and $\mathcal{Q}$ are particle/photon gain rates, all in the i-th segment's rest frame. More specifically,  `syn' refers to electron/positron synchrotron, `ics' to inverse Compton scattering, `inj' is particle injection, $\gamma\gamma$ is photon-photon pair production, `ssa' is synchrotron self-absorption, the electron escape time is simplistically assumed to be infinite for confined electrons so the `e,esc' term disappears, and the photon escape time is $t_{\rm \gamma,esc,i} = R_{\rm avg,i}/c$, and  $\mathcal{Q}_{\rm \gamma,n,i}^{\rm xseg}$ is the photon cross-segment contribution from each segment $n$, which can be upstream or downstream from segment $i$. 

Since the segments move at different speeds, in the $i$'s frame of reference each segment $n$ will be moving away from it at a relative speed $\beta_{\rm rel,n,i}$. Photons emitted from segment $n$ will reach segment $i$ from within a cone, but their average impact angle will be $0^\circ$, from one segment's center to the other. This simplifies our calculations for their Doppler factor, which will be $\delta_{\rm rel,n,i} = \Gamma_{\rm rel,n,i}^{-1}(1+\beta_{\rm rel,n,i})^{-1}$  (for an example, see Appendix~\ref{app:Gamma_rel}). All calculations are performed assuming that the particle population in each segment is in steady state and neglect the light travel time between segments. In other words, the jet has been operating long enough so that all segments are continuously illuminated by the emission of all other segments, which is in line with our iterative numerical method.
Therefore, if the steady-state photon number density in segment $n$ is $\left({\rm d}N/{\rm d}V {\rm d}\nu\right)_n$ and those photons are moving radially outwards at the rate they are being produced within segment $n$, then at a distance $z_{\rm n,i}$ that connects the centers of segments $i$ and $n$ that photon number density will be 
 \begin{equation}
 \left(\frac{{\rm d}N_\gamma}{{\rm d}V {\rm d}\nu}\right)_{\rm n,i} = \left(\frac{{\rm d}N_\gamma}{{\rm d}V {\rm d}\nu}\right)_{\rm n}\left(\frac{\Delta z_n}{2z_{\rm n,i}}\right)^2 \delta_{\rm rel,n,i}^2.
 \end{equation}

The code output from each segment is $\nu L_{\nu}$ computed in its rest frame (in units of $\mathrm{erg}/\mathrm{s}$), whereas the code inputs external photon fields as ${\rm d}N_\gamma/{\rm d}V {\rm d}\nu$ (in units of $\mathrm{cm}^{-3}\mathrm{Hz}^{-1}$). To convert the output into input for the next segment we calculate

\begin{equation}
\frac{{\rm d}N_\gamma}{{\rm d}V {\rm d}\nu} = \frac{\nu L_{\nu} R_{\rm avg}} {h \nu^2 c V},  
 \end{equation}
where $V'$ is the volume of the segment in the comoving frame (see Appendix~\ref{app:segments}). The photon field is considered isotropic in each segment.
Accordingly, 
 \begin{equation}
\mathcal{Q}_{\rm \gamma,n,i}^{\rm xseg} = \frac{c V_i}{R_{\rm avg,i}}\left(\frac{{\rm d}N_\gamma}{{\rm d}V {\rm d}\nu}\right)_{\rm n,i}.
 \end{equation}

\section{Results}
\label{sec:results}
In this section we present the electromagnetic signatures from striped blazar jets. In addition to the baseline model, we present results for alternate jets by varying one parameter at a time (see Table~\ref{tab:params}).

\subsection{Jet emission}
\begin{figure}
\includegraphics[width=0.45\textwidth]{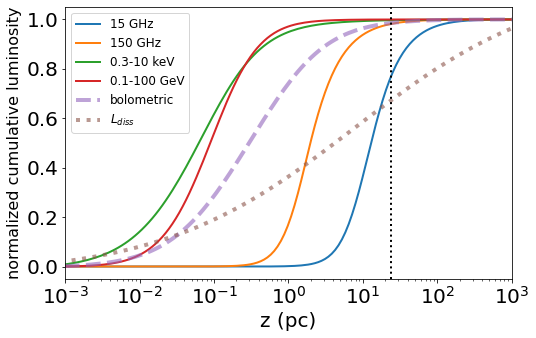}
\caption{Cumulative profiles of observer-frame photon luminosity in specific energy bands (solid lines), bolometric photon luminosity (dashed line), and dissipated power (dotted line) for the baseline model. Cumulative profiles are normalized to their respective maximum values. The vertical line indicates the distance $z_{\rm rec}/2$.}
\label{fig:cumulative}
\end{figure}

Although the entire jet can be modeled as a series of interacting segments, that does not mean that all segments are equally important when it comes to the radiative output of the jet. To illustrate this, we present the cumulative jet emission profile for the baseline model in Figure~\ref{fig:cumulative}. We also show for comparison the cumulative dissipation profile (dotted line), defined as $\int_{z_0}^z {\rm d} \bar{z} \frac{dL_{\rm diss}}{d\bar{z}}$. While the dissipation profile, $dL_{\rm diss}/d\ln z$, shows a broad peak at $z_{\rm rec}/2 \approx 24$~pc (see middle panel in Figure~\ref{fig:sigma_dLdz}), about 40\% of power is dissipated beyond this distance. In contrast, almost all photon luminosity (dashed line) is emitted by a distance of $z_{\rm rec}/2$, indicating a decrease in the particle radiative efficiency further downstream the jet; particles cool less efficiently because of the combined effect of weaker magnetic fields and lower average particle energy (see Figure~\ref{fig:model-params}). Focusing next on the emission profiles in different energy ranges, we see that the GeV $\gamma$-ray luminosity profile starts to increase at about ten times greater distance than the X-ray luminosity. While X-ray photons are emitted copiously at small distances through synchrotron radiation, where the magnetic field is stronger, GeV $\gamma$-ray photons are produced through synchrotron self-Compton, requiring the build up of sufficient number density of low-energy photons in the jet. The emission at low-frequencies, e.g. 15 GHz and 150 GHz, starts increasing beyond parsec scales, suggestive of the high opacity to synchrotron self-absorption at smaller distances.  
 
\begin{figure}
\includegraphics[width=0.45\textwidth]{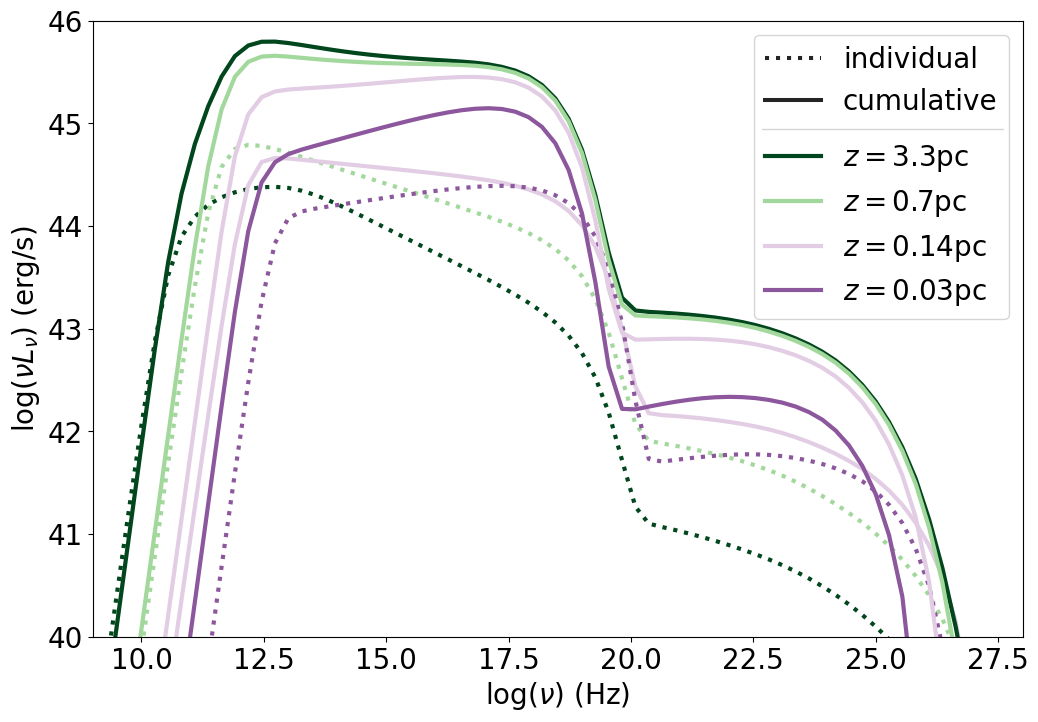}
\caption{Spectral energy distribution (SED) of various segments within a blazar jet with continuous magnetic dissipation, at different distances from the black hole. For each segment we plot the SED emitted from it (dotted lines) as well as the cumulative emission from all segments up to and including that one (solid lines).}
\label{fig:single_and_cumulative}
\end{figure}

In Figure~\ref{fig:single_and_cumulative}, using our baseline model we plot the SEDs emitted from four segments at around the distance where luminosity is expected to peak in the X-rays and $\gamma$-rays (green and red lines in Figure~\ref{fig:cumulative}). The emission from individual segments has different spectral shape and luminosity, reflecting differences in local plasma conditions that affect the properties of the radiating particles (see e.g. Fig.~\ref{fig:model-params}). Although the synchrotron peak of the spectrum from individual segments shifts to lower energies as their distance increases (dotted lines), the spectrum of the cumulative emission becomes flat (solid lines), as it includes contributions from a large number of segments.

\begin{figure*} 
\includegraphics[width = 0.49\textwidth]{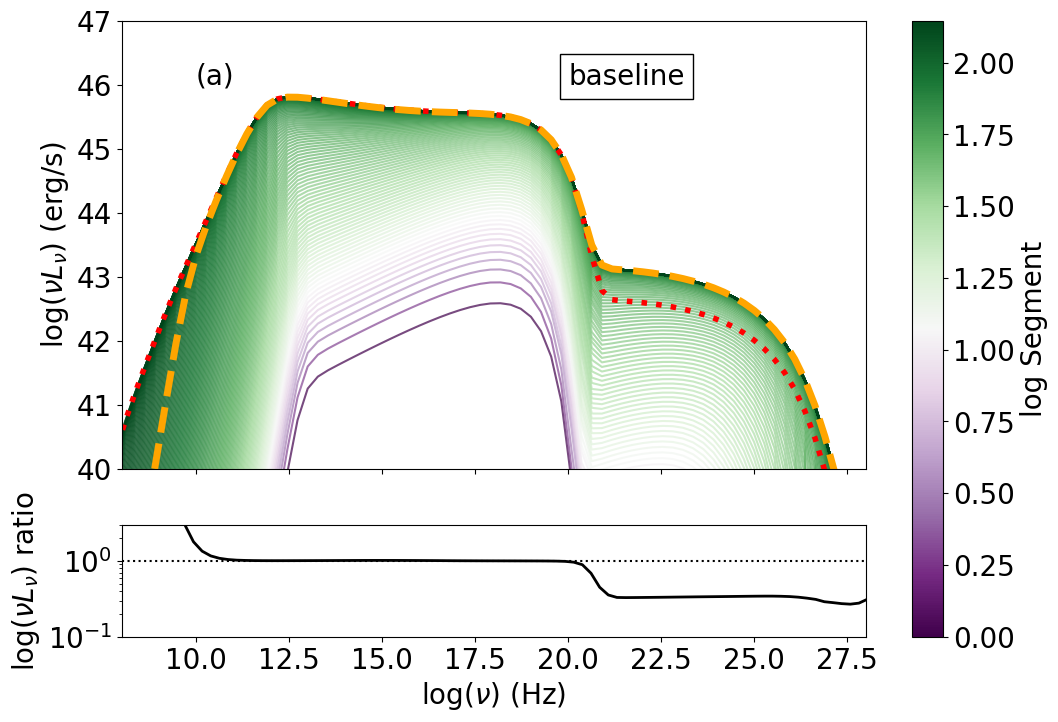}%
\includegraphics[width=0.49\textwidth]{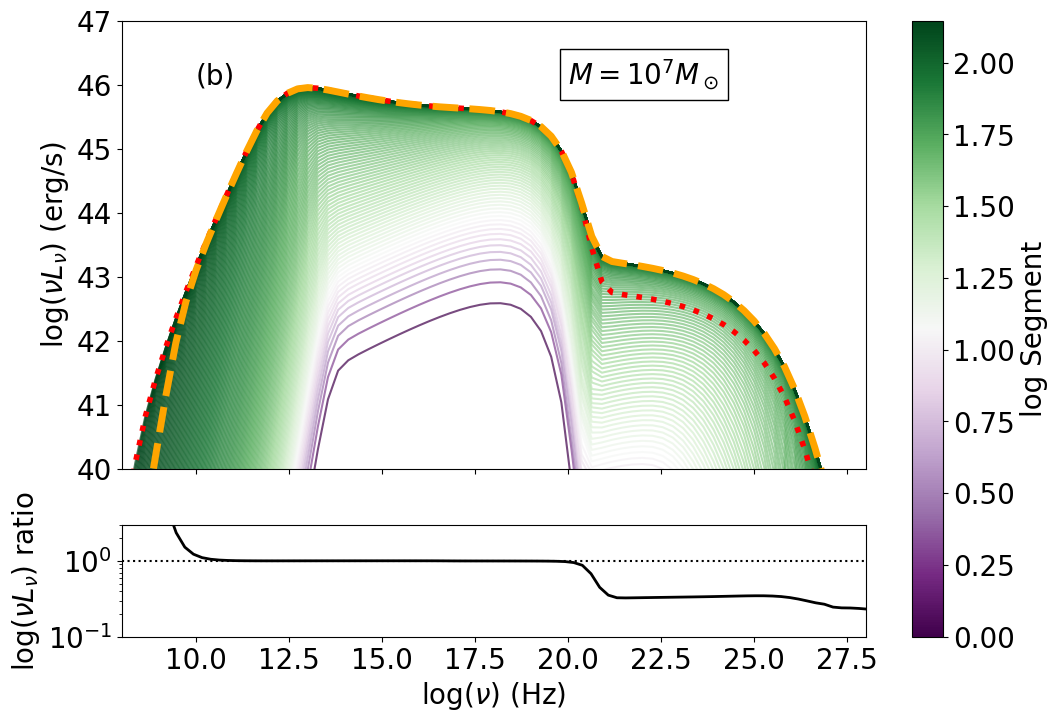}
\includegraphics[width = 0.49\textwidth]{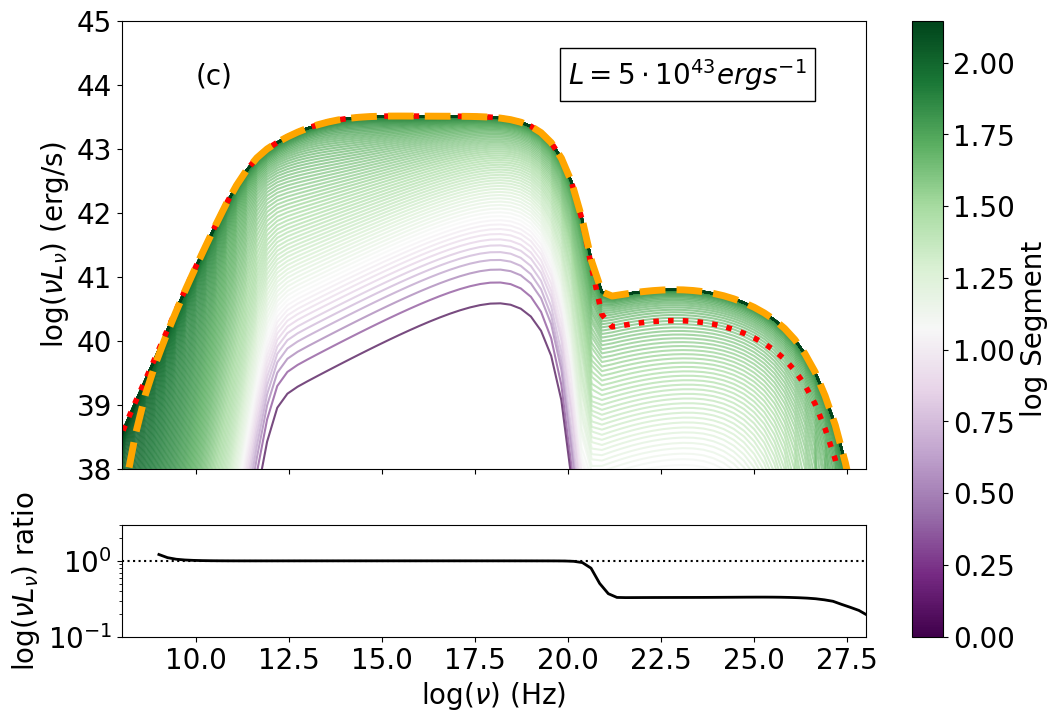}%
\includegraphics[width = 0.49\textwidth]{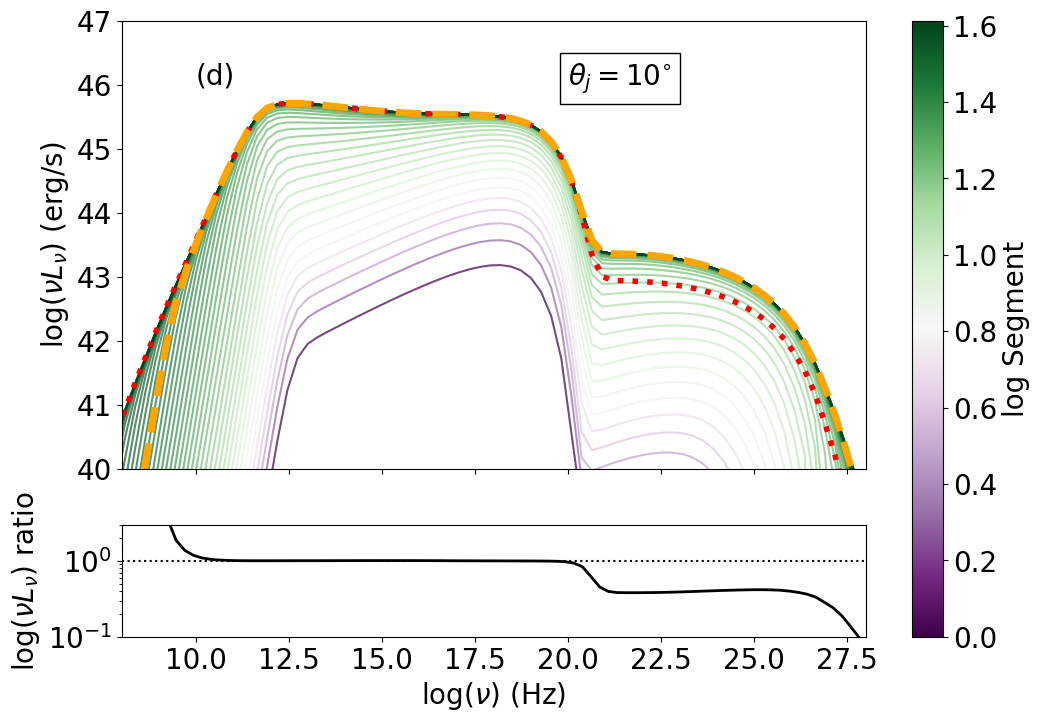}
\includegraphics[width = 0.49\textwidth]{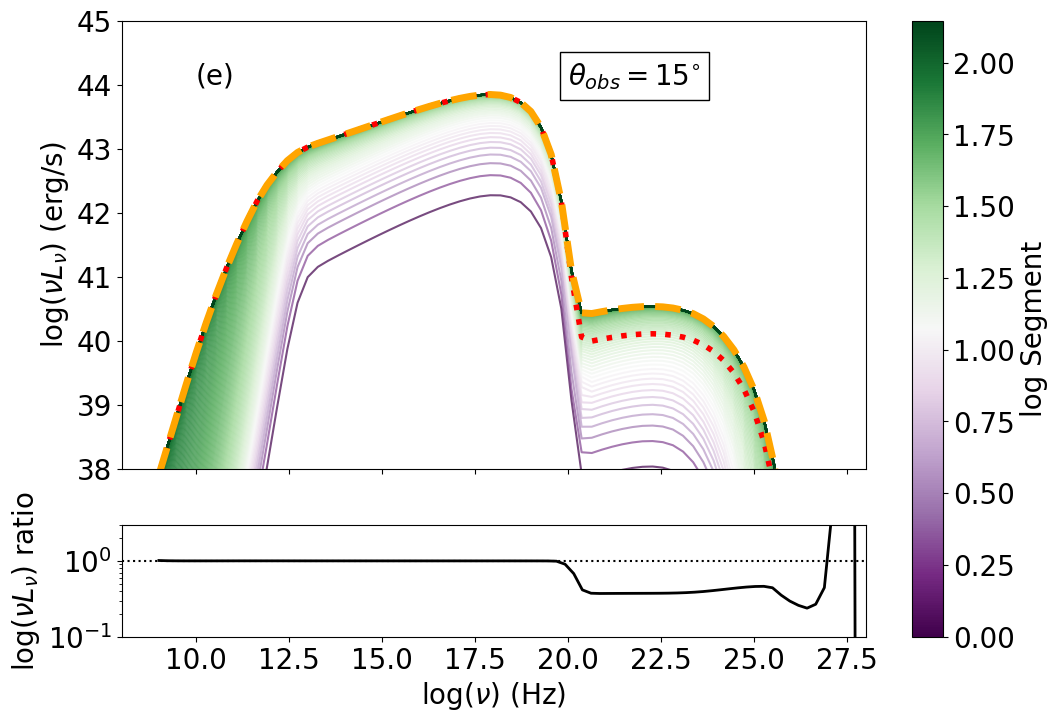}%
\includegraphics[width = 0.49\textwidth]{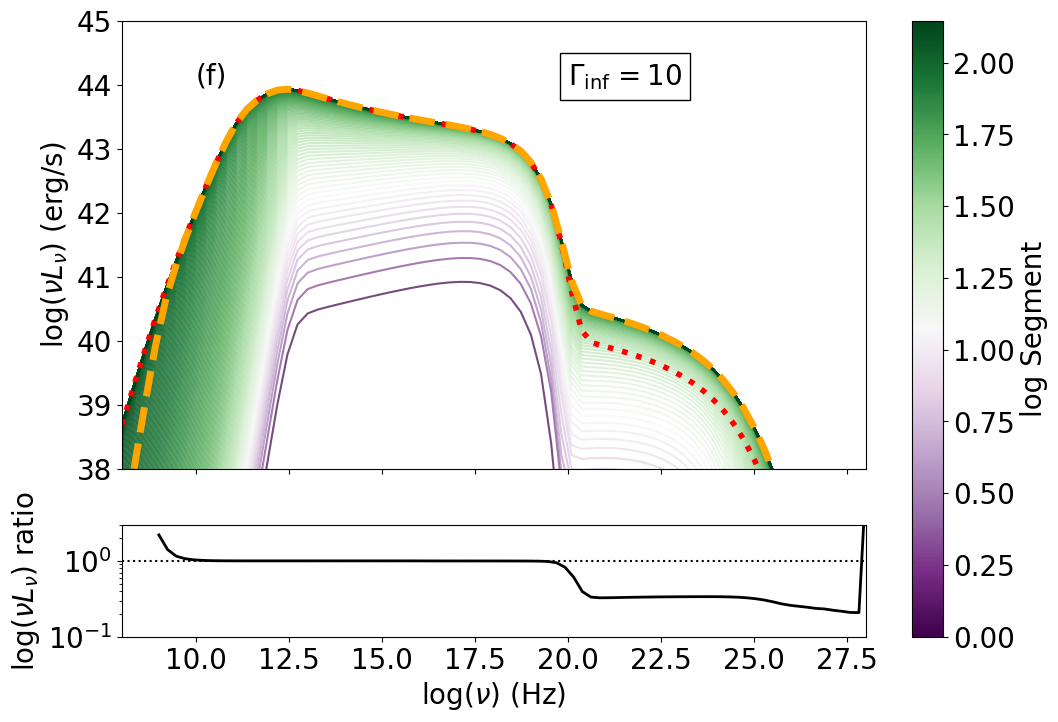}
\caption{
Spectral energy distribution (SED) of a blazar jet with continuous magnetic dissipation for various model parameters. Starting from the baseline model in panel (a), where the $z_{\rm rec}/2$ distance corresponds to segment No. 102, the other panels show models calculated by changing one parameter at a time: b) $M=10^7 M_\odot$, 
c) $L=5\cdot10^{43}$~erg s$^{-1}$, d) $\theta_{\rm j} = 10^\circ$, e) $\theta_{\rm obs} = 15^\circ$, and f)  $\Gamma_\infty = 10$. In each panel, spectra from emission up to each each segment along the jet are plotted with different colors. The dashed orange line indicates the spectrum arising from the distance where the dissipation profile peaks. The dotted red line shows jet spectrum calculated without taking into account photon spillage between segments. The ratio of these two spectra (in logarithmic scale) is displayed in the bottom panel of each plot.
}
\label{multi-sed}
\end{figure*}

Figure~\ref{multi-sed} presents the cumulative SEDs from jets with different properties, such as total power $L$, opening angle $\theta_{\rm j}$, and the asymptotic Lorentz factor $\Gamma_\infty$. We also show the results for the baseline model for a larger viewing angle $\theta_{\rm obs}$ and a smaller black hole mass. In each panel, colored curves show the spectrum produced in the jet up to each segment while going downstream (in total 140 segments were considered). In our scheme, after convergence has been achieved, the emergent spectrum from the outermost segment includes the contributions of all upstream segments, either directly (if photons escaped freely) or after being reprocessed by intermediate segments, as outlined in Fig.~\ref{fig:sketch}. We note that most of the photon emission, except the low frequencies ($\lesssim 15$~GHz), is already produced at $z_{\rm rec}/2$ (see dashed orange lines) in agreement with the cumulative emission profiles displayed in Fig.~\ref{fig:cumulative}. In each panel, we also show the photon spectrum computed by simply summing up the contributions of all segments (dotted red lines). This calculation neglects potential contributions from other segments to a segment's local photon density. We typically find that this approach underestimates the inverse Compton luminosity by a factor of approximately three. 

Inspection of the panels in Fig.~\ref{multi-sed} shows that the emergent jet emission has almost a universal SED, which is mostly determined by the local jet conditions at $z_{\rm rec}/2$, except for the low-frequency part of the spectrum which emerges from segments beyond $z_{\rm rec}/2$. Furthermore, the calculation of the SED using only local photon densities, as in standard one-zone models, would underestimate the ICS emission. The Compton ratio of the emergent spectrum, loosely defined as the ratio of luminosities of the ICS and synchrotron components, is $\sim 10^{-3}$ in all cases. This finding agrees with the results for the ratio $U'_{\rm B}/U'_{\rm e}$ (because all of the electron energy is radiated as synchrotron photons), which can be used as a proxy of the inverse of the Compton ratio; the minimum value of this ratio is $\sim 10^3$ and is attained close to $z_{\rm rec}/2$ as shown in Fig.~\ref{fig:model-params}. 

Even though several characteristic length scales of the problem ($l_{\min}$, $z_{\rm rec}$, and $z_0$) scale linearly with the black hole mass $M$, its impact on the final SED is small. The energy dissipation rate within each segment is independent of the black hole mass. However, for a smaller black hole mass, as in panel (b), each segment is located at smaller distances compared to the baseline model and has correspondingly smaller volume. 
Lowering the jet power, in panel (c) leads to a somewhat weaker low frequency emission compared to the rest of the spectrum. Increasing the jet opening angle, in panel (d) changes remarkably little, given how drastically the jet geometry is altered. The energy dissipation rate is independent of the jet's geometry, so a larger $\theta_{\rm j}$ results in fewer and larger segments with increasingly smaller energy densities. Given that the SEDs are synchrotron-dominated, the lower energy densities per segment are counterbalanced by their integration within the larger volume, to obtain luminosity. Therefore, wider jets with the same total power are expected to have the same SEDs. When the viewing angle is increased, on the other hand, in panel (e), the SED appears drastically different. The jet viewing angle $\theta_{\rm obs}$  strongly affects the Doppler boosting of the emitted radiation. 
The emergent spectrum is less luminous than all other spectra due to the lower Doppler factor, and has a rising synchrotron spectrum in $\nu L_{\nu}$ representation, similar to those emerging from the innermost regions of the jet where $\sigma \gg 1$ (see e.g. purple lines in Fig.~\ref{fig:single_and_cumulative}). For larger viewing angles, the location of dominant emission shifts upstream (i.e., closer to the jet base), because  beaming cones of segments located further downstream ``miss'' the observer.  Finally, panel (f) shows the effect of a much lower value of $\Gamma_{\infty}$. As before, the lower Doppler boosting results in lower observed emission overall, but this time the effect is more widespread along the jet, so its overall shape does not change much from the baseline case.

\subsection{External radiation fields}
In this section, we discuss if external radiation fields could enhance the ICS emission of our jet models. We consider first the radiation field from the broad line region (BLR). Its energy density in a segment's rest frame can be written as \citep{2009MNRAS.397..985G} 
\begin{equation}
U'_{\rm BLR}(z) = \frac{\xi_{\rm BLR} L_{\rm d}}{4\pi R_{\rm BLR}^2 c} f(\Gamma)
\label{eq:Ublr}
\end{equation}
where $R_{\rm BLR} \simeq 10^{17}~L_{\rm d, 45}^{1/2}~\rm cm$ and the Doppler boosting factor is 
\begin{equation}
f(\Gamma)  = \left \{ \begin{array}{ll}
                    \frac{17\Gamma^2}{12}, z < R_{\rm BLR} \\
                    \frac{\Gamma^2}{3\beta} \left( 2\left(1-\beta \mu_+ \right)^3 - (1-\beta \mu_-)^3 - (1-\beta)^3 \right), z > 3 R_{\rm BLR}
                      \end{array}
                       \right. 
\end{equation}
where $\mu_{\pm} = (1\pm R_{\rm BLR}^2/z^2)^{\mp 1/2}$. For distances $ R_{\rm BLR} \le z \le 3 R_{\rm BLR}$ we perform a linear interpolation of $\log_{10} U'_{\rm BLR}$, following \cite{2009MNRAS.397..985G}.

\begin{figure}
    \centering
    \includegraphics[width=0.9\linewidth]{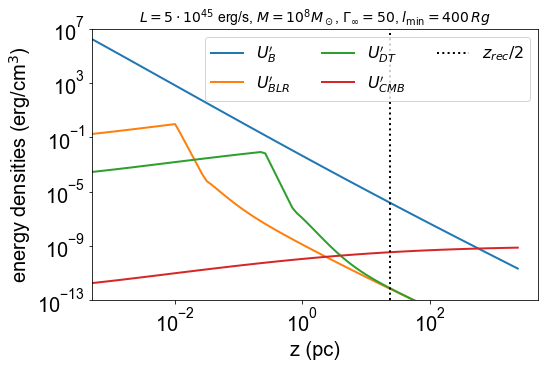}
    \caption{Energy densities of magnetic field and external radiation fields plotted as a function of distance along the jet for the baseline model. All energy densities are measured in the local rest frame of a segment located at distance $z$.}
    \label{fig:ext-phot}
\end{figure}

We search for conditions that lead to $z_* = z_{\rm rec}/2  \sim R_{\rm BLR}$ and  $U'_{\rm B} \lesssim U'_{\rm BLR}$ at the distance where the dissipation profile peaks. This condition is needed to channel the radiated energy to ICS more effectively than synchrotron radiation. Noting that $\Gamma \approx \Gamma_\infty$ and $\sigma \approx 1$ at $z_*$, we derive the following relations
\begin{eqnarray}
\label{rel1}
    L_{45} \approx 625 \, \beta_{\rm rec, -1}^{-2} \,  \lambda_2^2 \, M_8^2 \, \Gamma_{\infty, 1}^5 \\ 
    L_{45} \lesssim 0.3 \, \theta_{\rm j, -2}^{-2} \, \beta_{\rm rec, -1}^{-2} \,  \lambda_2^2 \, M_8^2 \, \Gamma_{\infty, 1}^8
    \label{rel2}
\end{eqnarray}
where we used Eq.~(\ref{eq:B}) and (\ref{eq:Ublr}) without the term in the parenthesis for simplicity. Moreover, $\lambda = l_{\min}/R_{\rm g}$ and we introduced the notation $q_X = q/10^X$ (the black hole mass is in units of $M_\odot$ and $\theta_{\rm j}$ is in radians). Both relations are simultaneously satisfied if $\Gamma_{\infty,1} \gtrsim 13\,  \theta_{\rm j, -2}^{2/3}$, which is extremely high for blazar jets \citep[e.g.,][]{2017MNRAS.468.4992P, 2019ApJ...874...43L}. Therefore, the external Compton scattering of BLR photons will be subdominant compared to the jet synchrotron emission for all relevant parameters. 

Although similar arguments can be made for the radiation field from the dusty torus (DT), the energy density of the cosmic microwave background radiation (CMBR) can surpass that of the magnetic field at kpc scales. To illustrate more clearly the above, we plot in Fig.~\ref{fig:ext-phot} the energy density of external radiation fields (BLR, DT\footnote{This is modeled according to Eq.~(\ref{eq:Ublr}) with $\xi_{\rm DT} = 0.1$  and $R_{\rm DT} = 2.5 \cdot 10^{18}~L_{\rm d, 45}^{1/2}$~cm.}, and CMBR) as measured in the local frame of the jet plasma at distance $z$ for the baseline model.

\subsection{Localized emission from large-scale jet}
In this section we focus on the emission arising from individual segments located at tens of parsecs. In particular, we
consider distances of 10, 100, and 1000~pc. While the contribution of other segments to the produced emission is taken into account during calculations as a source of external radiation in the segment, it is omitted from the displayed SED since it does not originate in that specific segment.

\begin{figure}
\includegraphics[width=0.45\textwidth]{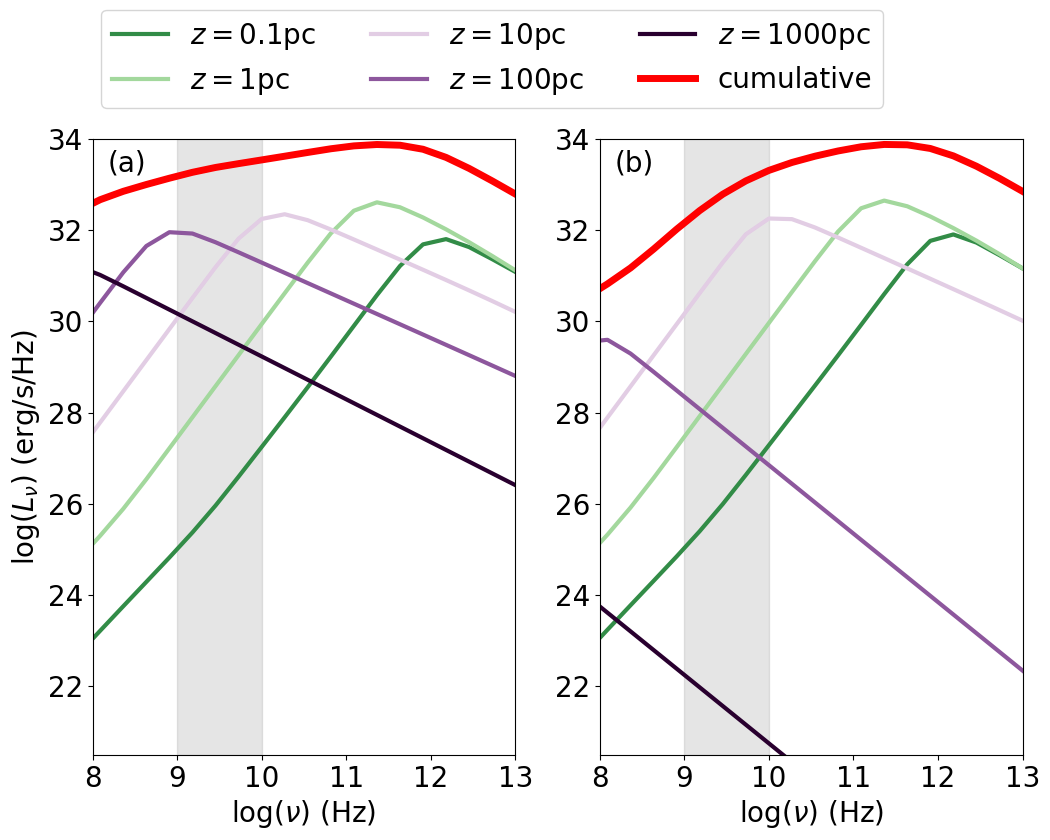}
\caption{SEDs from five specific segments (located at distances of 0.1 pc to 1000 pc) for (a) the baseline case where $a=1.4$ and for (b) $a=\infty$. We have focused on the low energy part of the spectrum and specifically in the range from 1~GHz to 10~GHz (shaded region).}
\label{fig:a1p4-inf}
\end{figure}

\begin{figure}
\includegraphics[width=0.45\textwidth]{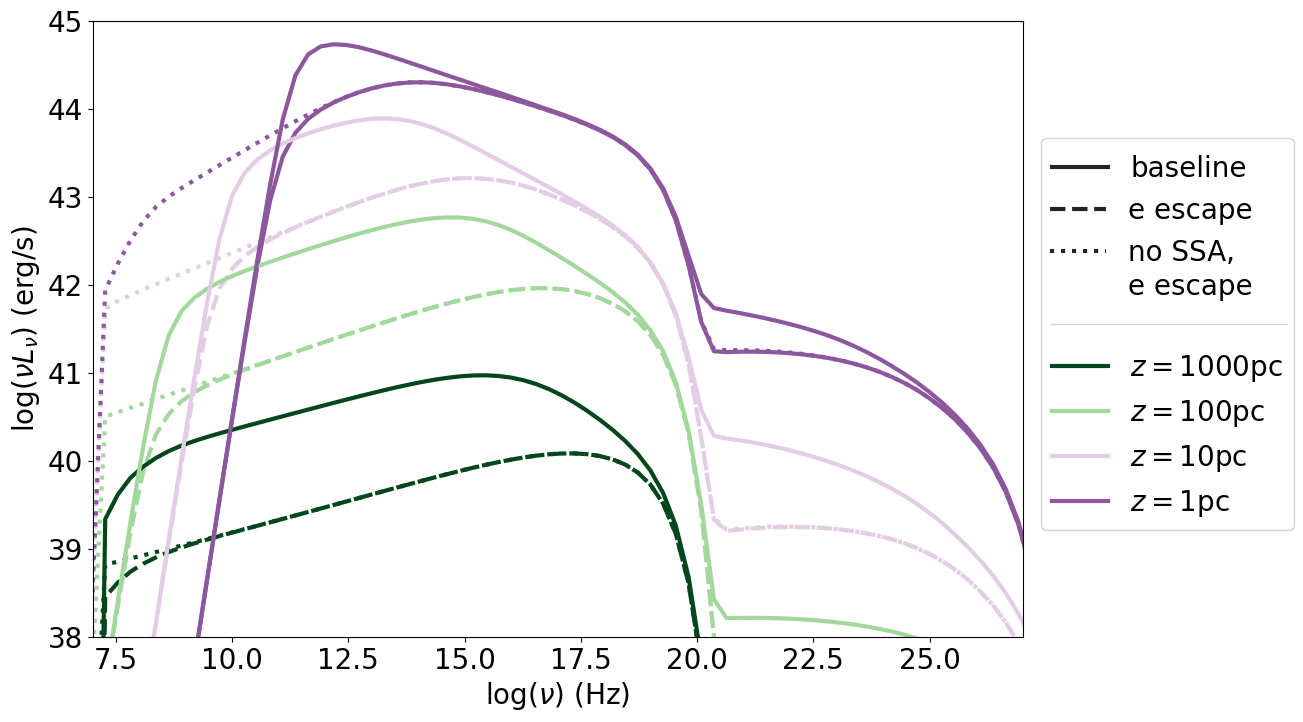}
\caption{SEDs from four specific segments (located at distances of 1 pc, 10 pc, 100 pc, and 1000 pc) for the baseline case where $a=1.4$ (solid lines)  and for the addition of electron escape (dashed line) and the removal of SSA losses (dotted lines).}
\label{fig:a1p4-3cases}
\end{figure}

In Figure~\ref{fig:a1p4-inf} we plot the low energy parts of the SEDs for the baseline model ($a=1.4$) and for $a=\infty$. 
The individual segment spectra don't exhibit the flatness, over a wide range of frequences, found by \cite{2021MNRAS.502.1145Z}; however, the cumulative emission (red lines), which at low energies is mostly affected by several segments clustered around some distance z, is very similar to what is shown by \cite{2021MNRAS.502.1145Z}.
In the latter's treatment of the striped jet model the $L_\nu$ spectra exhibited a broad spectral peak around 1-10 THz, whereas in our case the $L_\nu$ synchrotron peak sometimes extends to very low frequencies for individual segments, while it peaks at around 0.1-1THz in cumulative emission.
Next we estimate the radio spectral index $p$, defined as $L_{\nu}\propto \nu^p$, in the range 1~GHz $-$ 10 GHz, where such measurements are usually performed. The spectral indices
are listed in Table~\ref{tab:radio_slopes}, where we have also listed the case of $\alpha = 2$, not shown in Figure~\ref{fig:a1p4-inf}. 
At lower $\alpha$ more energy is dissipated further downstream as shown in Fig.~\ref{fig:sigma_dLdz}. The negative values of $p$ at large distances for the other values of $a$ suggest that the emission there is optically thin. This is illustrated in Fig.~\ref{fig:a1p4-3cases}. If we turn off synchrotron self absorption (SSA) then $p$ is very similar, and negative, from 1 to 1000 pc away from the black hole (dotted lines). Once SSA is accounted for (dashed lines), the value of $p$ in the frequency range 1~GHz - 10~GHz depends on the frequency for which SSA losses become important for the conditions of the jet at that distance. The SED itself peaks at ever higher frequencies downstream because that is regulated by the electron energy at which synchrotron losses overcome the electron escape timescale. Once electron escape is removed from the model (solid lines), as in our baseline case, there is an electron pileup at lower energies that causes the SED to skew to lower frequencies.

\begin{table}
\caption{Radio spectral index $p$ defined as $L_\nu \propto \nu^p$ between 1 GHz and 10 GHz}. 
\label{tab:radio_slopes}
\begin{center}
\begin{tabular}{l c c c}
\hline
 $z$ (pc) & \multicolumn{3}{c}{$p$ for:} \\
           & $\alpha = 1.4$ & $\alpha = 2$ & $\alpha = \infty$ \\ 
\hline 
10 & 2.42 & 2.41 & 2.41 \\
100 & -0.57	& -0.86	& -1.5 \\
1000 & -0.94 & -1.5 & -1.5\\
\hline
\end{tabular}
\end{center}
\end{table}

\begin{figure}
\includegraphics[width=0.45\textwidth]{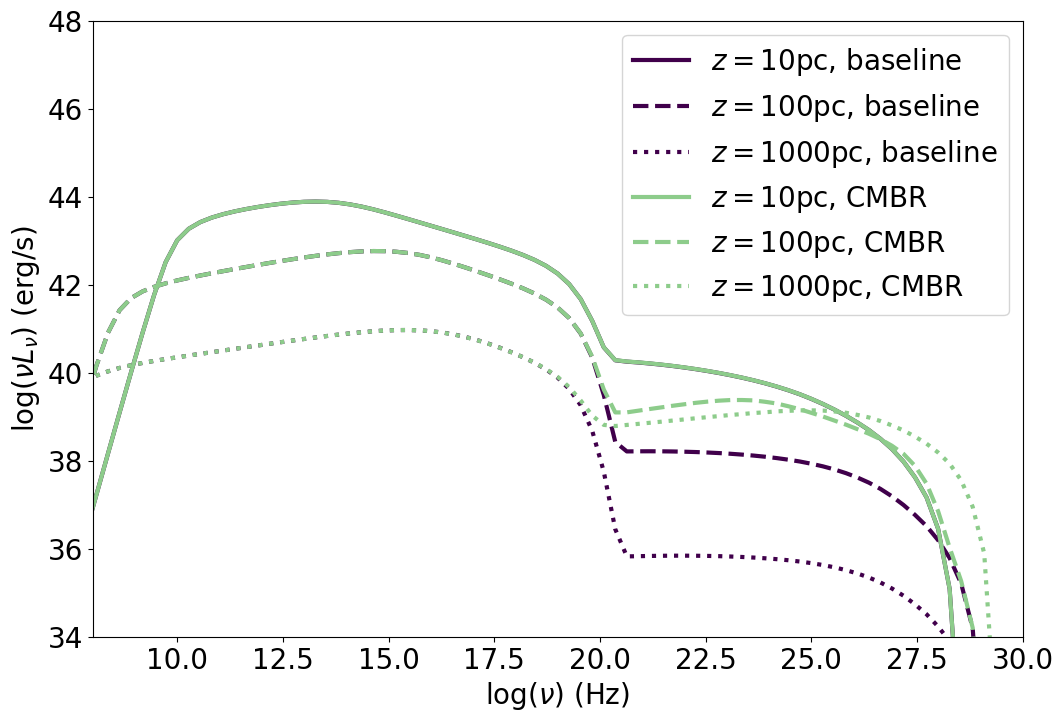}
\caption{SEDs from three specific segments (for distances of 10 pc, 100 pc, and 1 kpc) for  the baseline case (purple lines) and a model with external photon contribution from the CMBR (green lines). }
\label{fig:a1p4-cmbr}
\end{figure}

In Figure~\ref{fig:a1p4-cmbr} we plot localized emission from large distances (10 pc, 100 pc, and 1000 pc) but this time we compare the baseline model to a model with added CMBR external radiation field. As we expected from Fig.~\ref{fig:ext-phot} the external photons have no discernible effect at smaller distances, such as 10 pc, because their energy density in the segment frame is much lower than that of the magnetic field. At 100 pc we begin to see an effect at the inverse Compton hump, where there is an order of magnitude increase from the baseline, and at 1000 pc the increase is over three orders of magnitude from the baseline. Therefore, inverse Compton scattering off CMBR photons by electrons energized in the kpc scale jet may have a measurable effect on the TeV emission of large-scale jets, as has been suggested by \cite{meyer2015}.

\section{Discussion}\label{sec:discussion}
To calculate the total radiative or other particle emissions from an entire jet, one can discretize it into segments within each of which various parameters (such as the magnetic field strength) can be considered uniform. Typically, in each segment particle interactions are computed using local densities ~\citep[e.g.,][]{2019MNRAS.482.4798L, 2021MNRAS.500.2112K}. Such discretization, however, carries a risk of unaccounted for contamination; photons and charged massive particles from one segment may carry over to the next in non-negligible amounts, necessitating an iterative set of calculations to properly take such back and forth into account. In this work, we quantified the effects of photon spillage, working under the specific scenario of gradual magnetic dissipation. In all cases we explored, we found that the inverse Compton emission is enhanced by a factor of approximately three (in luminosity) when accounting for the cross-segment photons contributions. Photon spillage is expected to be more important in jets with higher photon densities, as those from GRBs.

In this model the jet is launched as Poynting-flux dominated outflow (having $\sigma_0 \gg 1$ at $z_0$). A possible mechanism for launching such outflows is the Blandford-Znajek process \citep{1977MNRAS.179..433B}. In this case, the total jet power can be written as $L_{\rm BZ}  = \kappa \Omega_{\rm H}^2 \Phi_{\rm H}^2 f_{\rm cor}/(4\pi c)$ where $\kappa \approx 0.05$ is related to the magnetic field geometry, $\Phi_{\rm H}=4\pi r_{\rm H}^2 B_{\rm H}$ is the magnetic flux threading the black hole horizon, and $B_{\rm H}$ is the magnetic field strength at the horizon \citep{2011MNRAS.418L..79T}. Here, $\Omega_{\rm H}=a_s c / (2r_{\rm H})$ is the rotational frequency of the horizon, $r_{\rm H}= R_{\rm g}(1+\sqrt{1-a_s^2})$ is the horizon radius,  $a_s$ is the black hole spin,  and $f_{\rm cor} = 1+1.38 \, (\Omega_{\rm H} R_{\rm g}/c)^2 - 9.2 \, (\Omega_{\rm H}R_{\rm g}/c)^4$ \citep{2011MNRAS.418L..79T}. One can obtain $B_{\rm H}$ requiring that $L = L_{\rm BZ}$ and compare with the magnetic field strength at the base of the outflow $B_0 \sim \Gamma_0 B'_0$, which can be derived from Eq.~\ref{eq:B} for $z=z_0$. Given that $z_0 \gg  r_{\rm H}$, we expect $B_{\rm H}> B_0$. If the opposite is found, then an energetic issue arises as a jet with luminosity as high as $L$ cannot be powered by this process. The ratio $B_{\rm H}/B_0$ can be written as
\begin{equation}
\frac{B_{\rm H}}{B_0} \approx \left( \frac{\beta_0}{8\pi \kappa a_s^2 f_{\rm cor}^2(a_s)}\right)^{1/2} \frac{z_0 \theta_{\rm j}}{R_{\rm g}} \Gamma_\infty^{-1/2} \approx 3 \left(\frac{\zeta}{100}\right) \left(\frac{\theta_{\rm j}}{0.1}\right)\left(\frac{10}{\Gamma_\infty}\right)^{1/2}
\end{equation}
where $\beta_0 = \sqrt{1-1/\Gamma_0^2} \sim 0.8$,  $\zeta = z_0/R_{\rm g} \gg 1$ in our model, $f_{\rm cor} \sim 1$ is assumed, and the first numerical factor is of order unity for $a_s = 0.5 - 0.99$. Interestingly, $B_{\rm H}/B_0$ is independent of $M, L$ and is greater than unity for $\Gamma_\infty \sim 10 - 50$ \citep{2019ApJ...874...43L} and $\theta_{\rm j} \sim 0.3/\Gamma_\infty$ that are typical for blazar jets \citep[e.g.,][]{2017MNRAS.468.4992P}.

A related study of reconnection-powered blazar emission was presented in \cite{stathopoulos2026}, where the radiative output of a localized dissipation region was explored as a function of its distance $z$ from the SMBH, while evolving the jet bulk properties (e.g. $\Gamma(z)$, $\sigma(z)$) and the external photon fields from the disk, BLR, and dusty torus. That framework was also extended to multi-messenger predictions by injecting comparable power into relativistic protons and pairs, which is a fraction $\varphi_{\rm rec}$ of the Poynting flux $L_{\rm P,jet}$ ($L'_{\rm i}\simeq \varphi_{\rm rec}L_{\rm P,jet}/(4 \Gamma ^2$)). Here, instead of a single reconnection zone, we model an extended striped jet with continuous dissipation over many decades in $z$, and we include radiative feedback via inter-segment photon exchange. The injected luminosity is connected to the properties of the stripes and the terminal Lorentz factor of the flow (see Eqs.~\ref{eq:Le}, \ref{eq:L_inj_nt}). In \cite{stathopoulos2026}, $\Gamma$ increases as $z^{1/2}$ up to a prescribed acceleration scale $z_{\rm acc}$, which is a free parameter, and then remains constant. The bulk acceleration profile is assumed, and the $z$-dependence of $\sigma$ is then obtained from it using the conservation of $\mu\equiv \Gamma(1+\sigma)$. In this work instead, a shallower profile of the jet bulk Lorentz factor is obtained ($\Gamma \propto z^{1/3}$) through the power dissipation profile, see Eqs.~(\ref{eq:dLdz}) and (\ref{eq:sigma}). These differences in the magnetization profiles are also mapped to the magnetic energy density profiles, with the latter decreasing faster with distance in the model of \cite{stathopoulos2026}. Both approaches predict distance-dependent emission regimes. Synchrotron/SSC dominates at high $\sigma$ close to the SMBH, whereas external Compton can become important near the BLR when the external photon energy density is boosted in the comoving frame. However, a key difference in the two studies is the predicted relation of $U'_B$ and $U'_{\rm ext}$. In \cite{stathopoulos2026}, as the flow approaches $\Gamma_{\infty}$ near the BLR and $U'_{\rm B} \lesssim U'_{\rm ext}$, external Compton can become prominent. In our model, $\sigma$ declines more gradually ($\sigma \sim z^{-1/3}$) and dissipation peaks around $z\sim z_{\rm rec}/2$ (typically at tens of parsecs, see Fig.~\ref{fig:ext-phot}), where BLR/disk fields are effectively de-boosted, so $U'_B \gtrsim U'_{\rm ext}$ in the segments dominating the bolometric output, yielding synchrotron-dominated SEDs. Finally, both models predict softer injected particle spectra at larger distances as $\sigma$ decreases. In our case, however, particle distributions with power-law slopes $s\gtrsim 3$ are obtained beyond $z_{\rm rec}$, at tens of parsecs from the central engine (see e.g. Fig.~\ref{fig:model-params}). On the contrary, such soft distributions could be achieved at sub-parsec scales in the model of \cite{stathopoulos2026} due to the faster decline of $\sigma$ with distance.

{Another series of similar studies were undertaken by \cite{Potter2012, Potter2012II, potter2012III}. There, they first treated the jet as a conical outflow of constant $\Gamma$ where $B$ changed with distance to conserve the magnetic flux going through each slice of a continuous jet \citep{Potter2012} and then they modeled it as a parabolic accelerating section followed by a conical decelerating section \citep{Potter2012II}, where $\Gamma$ increased to a maximum value before decreasing again. Particle acceleration was inferred by assuming that the number of electrons traveling across each slice of the jet was conserved while the total energy available for them increased. Slices further along the parabolic part of the jet had a higher bulk velocity, and that translated to higher speeds for the isotropized electrons within them. In the conical, decelerating, part of the jet, electrons were assumed to be accelerated at shocks. Their distribution was, throughout the jet, taken to be a power law of constant slope. That modeling, which also accounted for several sources of external photons, allowed for the emergence of Compton-dominated blazar emission. 

More recently, \cite{2021MNRAS.502.1145Z}  modeled the emission from a continuously accelerating jet, this time under the specific assumption of a striped jet where energy dissipation happens by magnetic reconnection, as in our work. They also obtained Compton-dominated spectra, but the discrepancy with our results can be explained by a methodological difference between their approach and ours. In Eq.~\ref{eq:Le} we calculated the dissipated power injected into non-thermal electrons as an integral of the differential energy dissipation rate within that segment exclusively. \cite{2021MNRAS.502.1145Z} instead calculated the dissipated power at distance $z$ as $P_{\rm diss}=z \, {\rm d}L_{\rm kin}/{\rm d}z$, in effect assuming that all dissipated power up until distance $z$ is available to the electrons at $z$. \cite{Potter2012II}, although following a different formulation, essentially make a very similar assumption. While it seems reasonable, under conditions of very gradual energy dissipation any energy transferred to non-thermal electrons is radiated away in timescales short enough to essentially remove it from the system. We tested energy retention through electrons that survive from segment to segment up to six times, in panel (c) of Fig.~\ref{multi-sed2}, and the Compton component of the SED increased compared to the synchrotron component, but not enough to match the energy density of the synchrotron photons in any segment.

\section{Conclusions}\label{sec:conclusions}

We have applied a model of gradual magnetic dissipation to a segmented blazar jet, with interaction of emitted photons between its segments. Inter-segment radiation spillage produces some power transfer effects, by a factor of three, in the part of the spectrum affected by inverse Compton scattering. However, in all our test cases that component is 
by far less luminous than the synchrotron component, 
which makes this physical scenario unlikely for powering the variety of observed blazar spectra. The typically considered external photon sources (BLR, dusty torus, CMB) are insufficient to make up for the overwhelming $U'_B$ dominance observed at all distances along the jet, so more specific scenarios may have to be explored to make this model viable for blazars. If the energization of particles is not gradual across the jet but peaks at shocks or other areas where the jet turns unstable then the observed emission may be quite different from what was presented here. The numerical framework we developed can be easily adapted to other theoretical scenarios for dissipation, and we provide it as an open-source code.

\begin{acknowledgements}
The authors acknowledge support from the Hellenic Foundation for Research and Innovation (H.F.R.I.) under the ``2nd call for H.F.R.I. Research Projects to support Faculty members and Researchers'' through the project UNTRAPHOB (Project ID 3013), which was active during the authors' research. D. G. acknowledges support from the NSF AST-2308090 and AST-2510569 grants.  
\end{acknowledgements}

%
%

\bibliographystyle{aa} 
\bibliography{A38.bib}

\appendix
\section{Numerical aspects}\label{app:numerical}
In this section we discuss in more detail some numerical aspects of our continuous jet model. 

\subsection{Derivation of segment sizes}
\label{app:segments}
For the purposes of denoting the segment locations and dimensions we can initially assume that each segment in the jet is spherical, but in which frame of reference? There are two obvious options, either the comoving frame or the observer's frame. The case of the comoving frame matches the methodology of typical single-region modelling. But requiring each sphere to be adjacent to the ones next to it, with no overlap, quickly runs into some problems. For simplicity, let us consider a conical jet with a single bulk Lorentz factor $\Gamma$. Since the jet opening angle, $\theta_{\rm j}$, is constant in the observer's frame, in the comoving frame it will be 
\begin{equation}
\tan \theta'_{\rm j} = \frac{\sin\theta_{\rm j}}{\Gamma (\cos \theta_{\rm j} -\beta)}.
\end{equation}
For moderate values of $\theta_{\rm j}$ (e.g. $3^\circ$) similarly moderate values of $\Gamma$ (e.g. 10) result in $\theta'_{\rm j} > 1$. When that happens, it is geometrically impossible to inscribe another sphere, in the comoving frame, adjacent to a previous one. Moreover, in the gradual dissipation model, there is no unique jet-comoving frame, because of the gradual bulk acceleration.

For these reasons, we elected to use spherical segments in the observer's frame as illustrated in Figure~\ref{fig:jet_pancakes}. With a constant $\theta_{\rm j}$, for each segment starting at distance $z$ its diameter/height will be 
\begin{equation}
\Delta z = \frac{2z\tan \theta_{\rm j}}{1-\tan \theta_{\rm j}}.
\label{eq:dz}
\end{equation}
In the comoving frame of a segment with bulk Lorentz factor $\Gamma$, the respective height will be $\Delta z' = \Delta z / \Gamma$ as illustrated in Figure~\ref{fig:jet_pancakes}.

\begin{figure}
\centering
\includegraphics[width = 0.45\textwidth]
{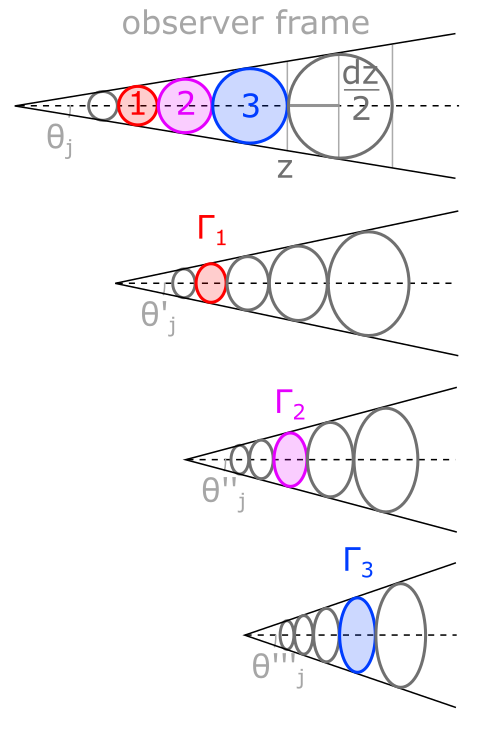}
\caption{Schematic of the jet as it appears from the reference frame of each segment (with examples here numbered 1, 2, 3 and coloured for clarity) as the Lorentz factor increases downstream. In the observer's frame the opening angle is $\theta_j$ and a spherical segment beginning at distance z will have a diameter dz. In each segment's comoving frame the spheres appear like increasingly pancake-like spheroids.
}
\label{fig:jet_pancakes}
\end{figure}

\begin{figure}
\centering
\includegraphics[width = 0.45\textwidth]
{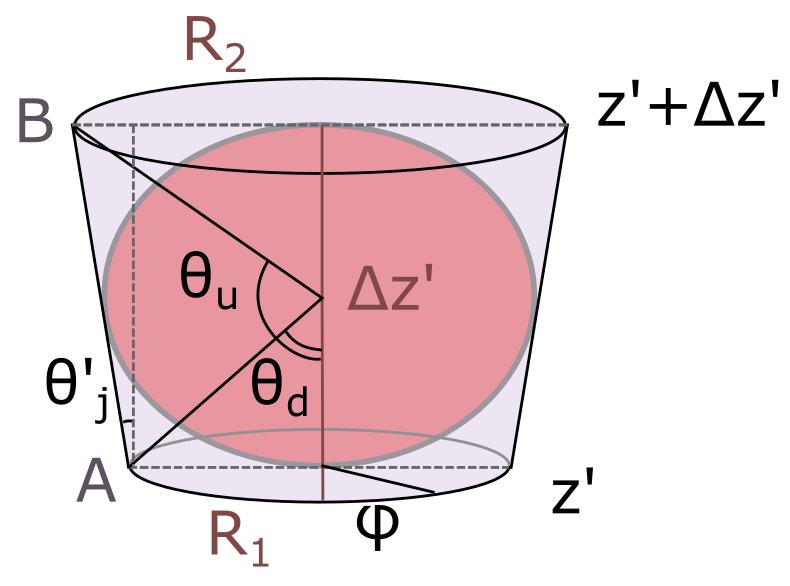}
\caption{Schematic of a jet segment. The part of the jet that inscribes the spheroid is a frustum.
}
\label{fig:frustum}
\end{figure}

\begin{figure}
\centering
\includegraphics[width = 0.45\textwidth]
{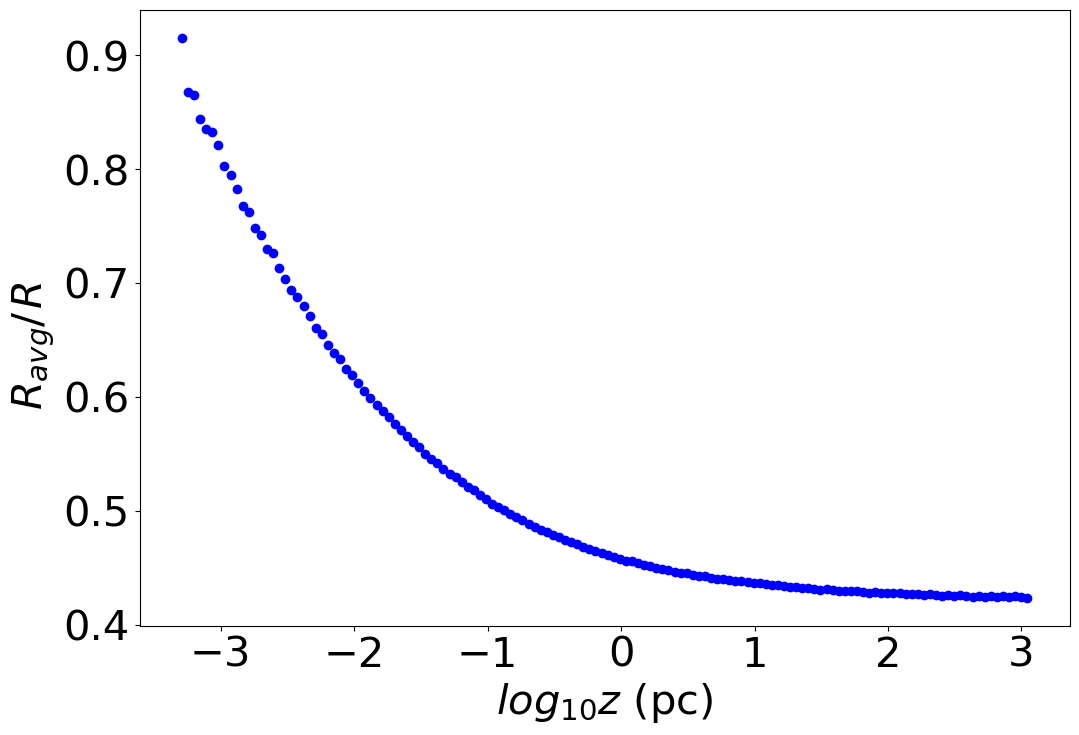}
\caption{Ratio of each segment's average radius $R_{\rm avg}$ over the cross sectional radius $R=\Delta z/2$ along the jet.
}
\label{fig:Ravg}
\end{figure}

The volume of each segment, in its comoving frame, can be simplistically calculated as a spheroid,
\begin{equation}
V' = \frac{4}{3} \pi \frac{\Delta z^2 \Delta z'}{8}
\label{eq:Vs}
\end{equation}
or a frustum, which is what we opted for in the numerical jet model (Figure~\ref{fig:frustum}):
\begin{equation}
 V' = \frac{\pi (\tan \theta_j)^2}{3} ((z+\Delta z)^2-z^2)z'.
\label{eq:Vs}
\end{equation}

For calculations of the escape time of photons (and in occasion of electrons) from each segment we will use the average distance from the center of the frustum to its edges. That is given by
\begin{eqnarray}
R_{\rm avg} = \frac{1}{4\pi} \left( \int_{0}^{\theta_d}\int_{0}^{2\pi}  \frac{\Delta z'/2}{\cos{\theta}} \sin\theta d\theta d\phi +  \right. \nonumber \\
\int_{\theta_d}^{\theta_u}\int_{0}^{2\pi} \frac{\sin(\theta'_j+\theta_d)\sqrt{R_1^2+\Delta z'/2}}{\sin{(\pi-\theta'_j-\theta)}} \sin\theta d\theta d\phi +  \nonumber \\
\left.\int_{\theta_u}^{\pi}\int_{0}^{2\pi} \frac{\Delta z'/2}{\cos{(\pi- \theta)}} \sin\theta d\theta d\phi \right), 
\end{eqnarray}
where $\theta_d$ and $\theta_u$ are the angles to points A and B along the frustum's edge (Figure~\ref{fig:frustum}). We compare $R_{\rm avg}$ to the segment's cross sectional radius $R = \Delta z/2$ along the jet in Figure~\ref{fig:Ravg}. We can see that if one were to use $R$ the escape timescale would be underestimated by a value of over two beyond the parsec scale, where the segments are more pancake-like than spherical. $R_{\rm avg}$ avoids that pitfall.

\subsection{Relative motion of segments} \label{app:Gamma_rel}
In the gradual dissipation jet model each segment has different Lorentz factor due to the continuous bulk acceleration of the jet plasma. 

We compute the relative Lorentz factor $\Gamma_{\rm rel}$ of the segments $i$ and $j$ in the jet. Their relative velocity is $\beta_{\rm rel} = (\beta_{\rm i} - \beta_{\rm j})/(1 - \beta_{\rm i} \beta_{\rm j})$ and
\begin{equation}
 \Gamma_{\rm rel}=(1-\beta_{\rm rel}^2)^{-1/2} = \Gamma_{\rm i} \Gamma_{\rm j} (1-\beta_{\rm i} \beta_{\rm j}).
\end{equation}
Our results are displayed in Figure~\ref{Grel} for some indicative cases. For the first segment (magenta line) there are no segments at smaller distances, $z$, along the jet axis, so $\Gamma_{\rm rel}$ increases with distance. For the last segment (blue line) the sequence of $\Gamma_{\rm rel}$ is reversed, since the segments closer to the beginning of the jet appear to move faster relative to the last one. Segments in the middle experience a milder increase in $\Gamma_{\rm rel}$ with other segments in both directions. We list the jet parameters and the first and last values of $z$ and $\Gamma$ in Table~\ref{tab-par-Grel}. Note that the maximum $z$, and the corresponding number of segments, are higher than those used in the other figures,  merely to better illustrate the differences between distant segments.

{\tt LeHaMoC} outputs $\nu L_{\nu}$, in units of $\mathrm{erg}/\mathrm{s}$ and inputs external photon fields as $dN/dV d\nu$, in units of $\mathrm{cm}^{-3}~\mathrm{Hz}^{-1}$. To convert the output of one segment, at distance $z$, into input for another segment, at distance $z_2$,
moving with a Doppler factor $\delta_{rel}$ relative to the first segment, we first calculate the number density of photons in the first segment's frame
 \begin{equation}
 {dN' \over dV' d\nu'} = { \nu' L'_{\nu} R'_{avg} \over  h \nu'^2 c V'},  
 \end{equation}
 and then convert it to the density present at the other segment (using double primes to denote the other segment's reference frame), which is at distance $z_2$,
 \begin{equation}
 {dN'' \over dV'' d\nu''} =  {dN' \over dV' d\nu'}\left({\Gamma dz''/2 \over z_2+dz_2/2 -z-dz/2}\right)^2 \delta_{rel}^2 a_{gg},  
 \end{equation}
 where $a_{gg}$ is the cumulative $\gamma-\gamma$ absorption rate from the intervening medium.

\begin{figure}
\centering
\resizebox{\hsize}{!}{\includegraphics[clip=true]
{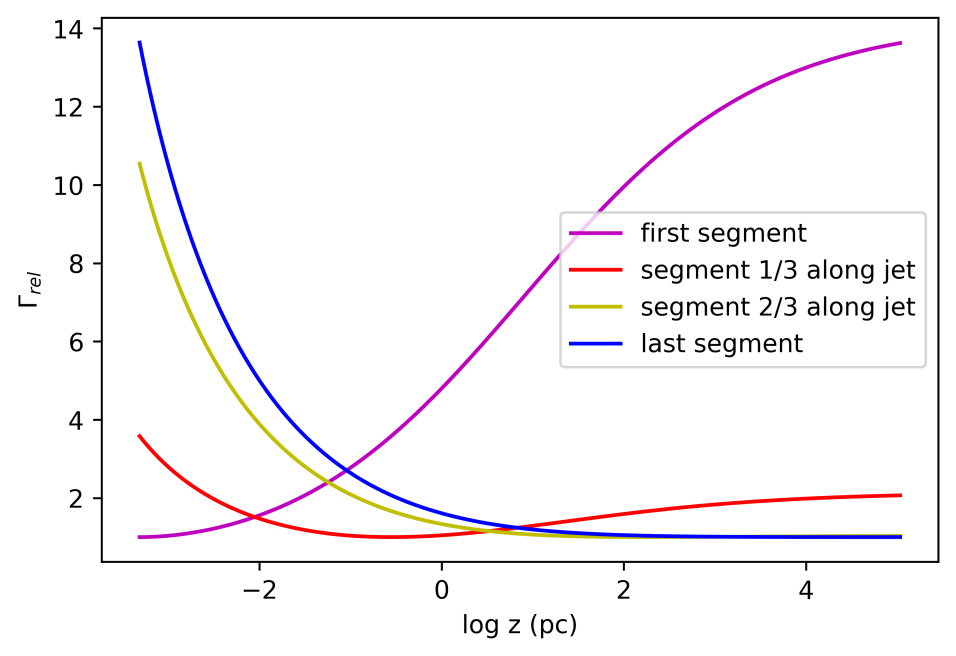}}
\caption{Relative Lorentz factor ($\Gamma_{\rm rel}$) between one segment and each other along the jet, until a certain distance where $\Gamma$ approaches $\Gamma_\infty$. The parameters used are listed in Table~\ref{tab-par-Grel}. 
}
\label{Grel}
\end{figure}

\begin{table}
\caption{List of parameters used for Figure~\ref{Grel}.}
\label{tab-par-Grel}
\begin{center}
\begin{tabular}{l c c}
\hline
 Parameter & Value\\
\hline
$\theta_j$ & $3^{\circ}$\\
$\Gamma_\infty$ & 50 \\
$L$ & $5 \cdot 10^{45}~\mathrm{erg/s}$ \\
$a$ & 1.4\\
segments & 197 \\
$z_1$ & $5\cdot10^{-4}$~pc\\
$z_{197}$ & $1.2\cdot10^{5}$~pc\\
$\Gamma_1$ & 1.9 \\
$\Gamma_{197}$ & 48 \\
\hline
\end{tabular}
\end{center}
\end{table}

\subsection{Number of iterations}
\label{app:iterations}
The number of iterations needed in our model to reach convergence is subjective, and depends on the required level of accuracy. In Figure~\ref{iterations} we show the total luminosity of the second hump, assumed to start at $\nu=10^{20} \mathrm{Hz}$, which is the part of the spectrum that changes appreciably with each iteration of photon spillage. Since odd-numbered iterations follow the segments upstream, only the even-numbered ones will show us changes in the jet's total emission. If we assume that Iteration~16 is representative of the jet's final state (the increase in luminosity from Iteration~14 is only 0.004\%), then we can use that as a benchmark to determine our desired accuracy. Iteration~4 produces 95\% of the luminosity of Iteration~16. Iteration~6 produces 99\%. Iteration~8 produces 99.7\%. For the illustrative purposes of this study 95\% was deemed sufficient, so we show results for Iteration 4 throughout this manuscript.

\begin{figure}
\centering
\resizebox{\hsize}{!}{\includegraphics[clip=true]
{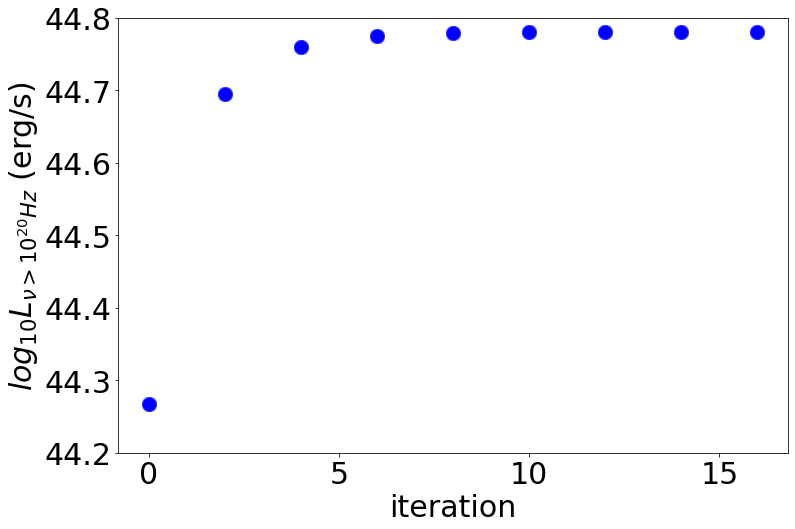}}
\caption{Integrated photon luminosity for $\nu>10^{20}$~Hz, where the ICS component is expected to emerge, for an increasing number of iterations, using the baseline values for all free parameters. 
}
\label{iterations}
\end{figure}

\section{Effects of parameters on jet emission}\label{app:SEDs}

In Fig.~\ref{multi-sed2} we present additional results obtained by altering more physical parameters or changing some aspects of the numerical treatment. Panel (a) shows electron escape. Its main difference to the baseline case is a flattening of the emission at low energies, since low energy electrons don't accumulate as much as before. Panel (b) uses a shorter acceleration timescale, with $\eta_{\rm acc}=100$ instead of $10^4$. This increases $\gamma'_{rad}$ and by extension $\gamma_{max}$ and the frequency of synchrotron emission. Panel (c) retains the electron distribution of the previous six segments (upstream) to better simulate the presence of previously accelerated electrons throughout the jet. This leads to a higher inverse Compton emission compared to the baseline model. In panel (d) the segment width has been reduced to 1/3 of the baseline's, making them more pancake-like. While this technical detail oughtn't affect the result, in practice we do see a difference, which may hint at an imperfection of the implementation of $R_{avg}.$ Panel (e) uses a higher jet power; the result is essentially shifted up from the baseline by the same factor. Panel (f) tests a reduced minimum stripe width, which results in more emission close to the black hole but not much of a difference at large scales.

\begin{figure*} 
\includegraphics[width = 0.49\textwidth]{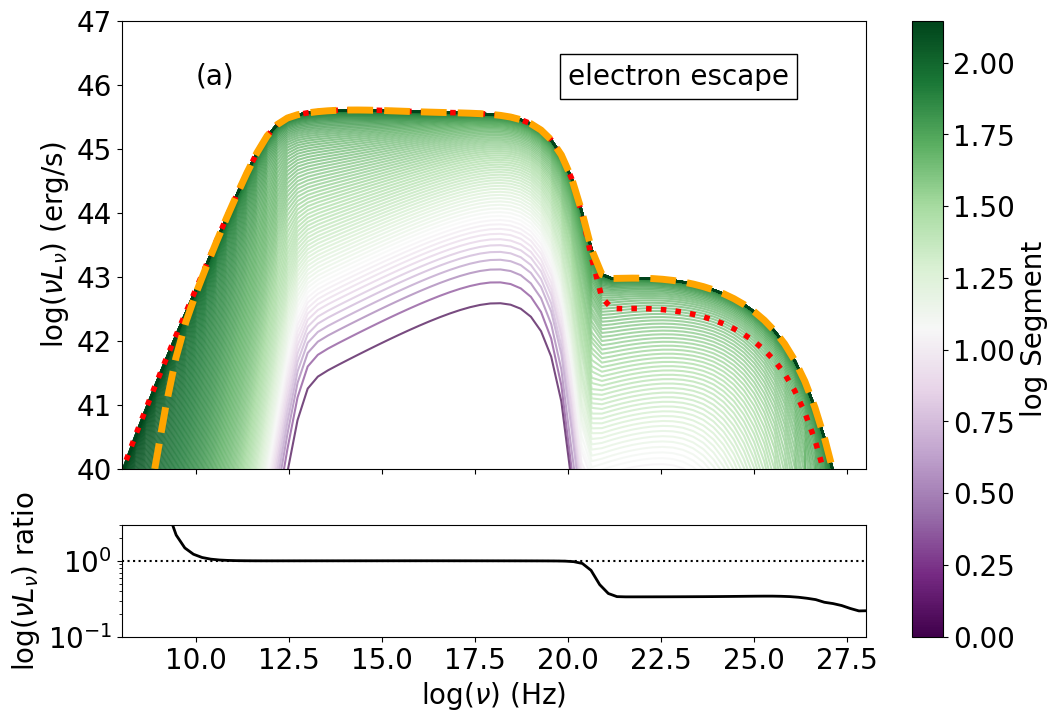}%
\includegraphics[width = 0.49\textwidth]{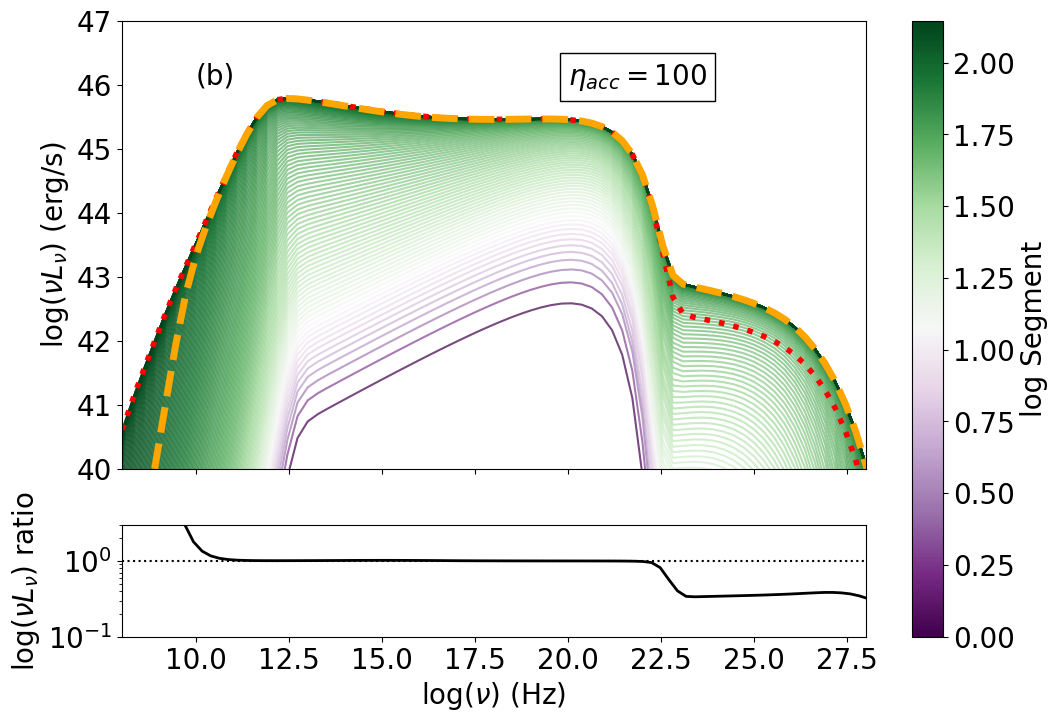}
\includegraphics[width = 0.49\textwidth]{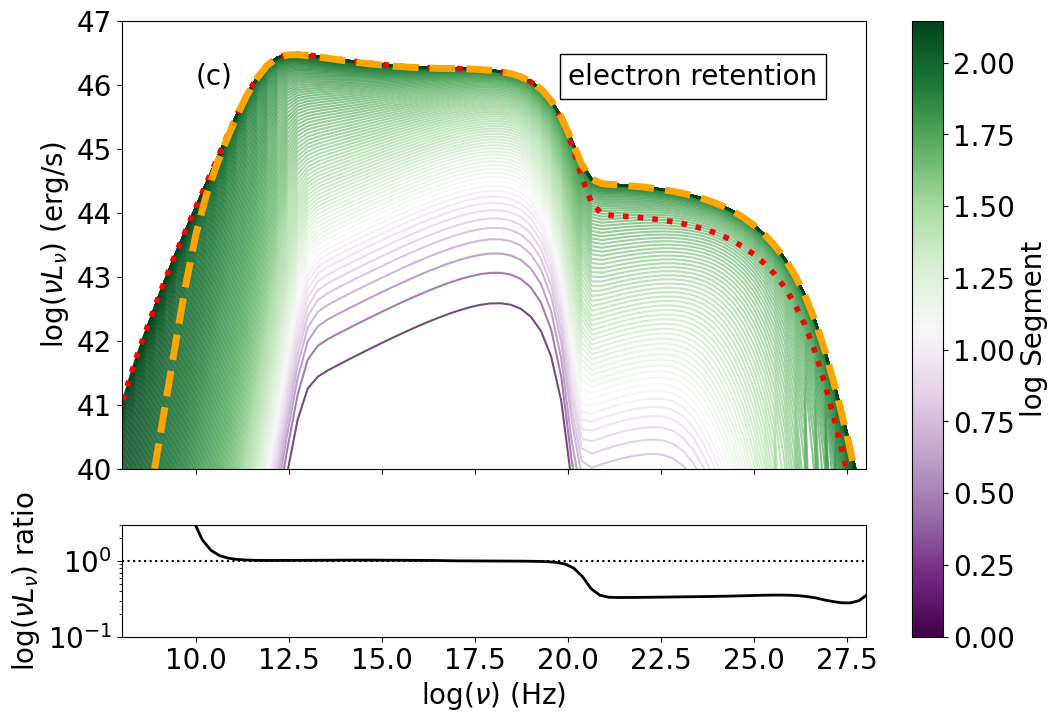}%
\includegraphics[width = 0.49\textwidth]{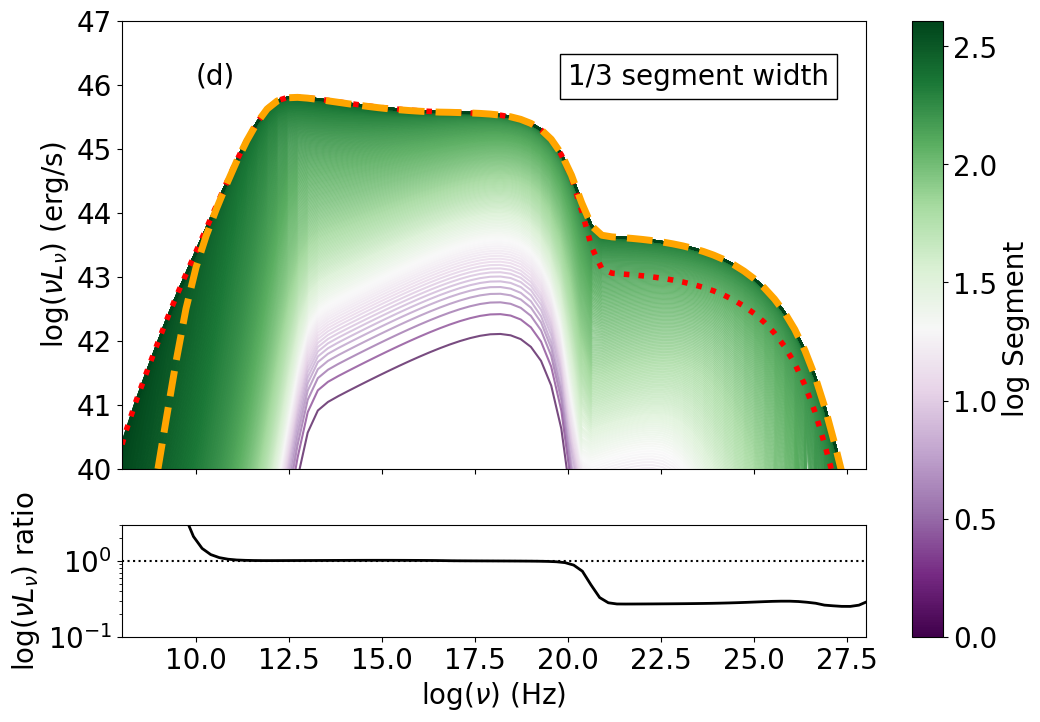}
\includegraphics[width = 0.49\textwidth]{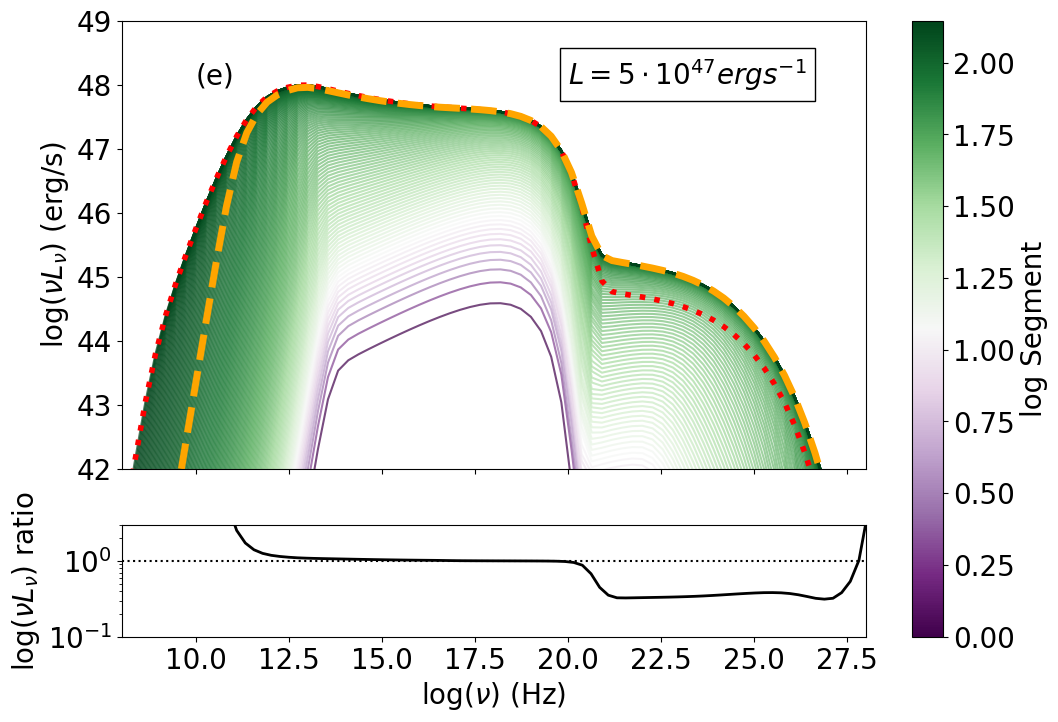}
\includegraphics[width = 0.49\textwidth]{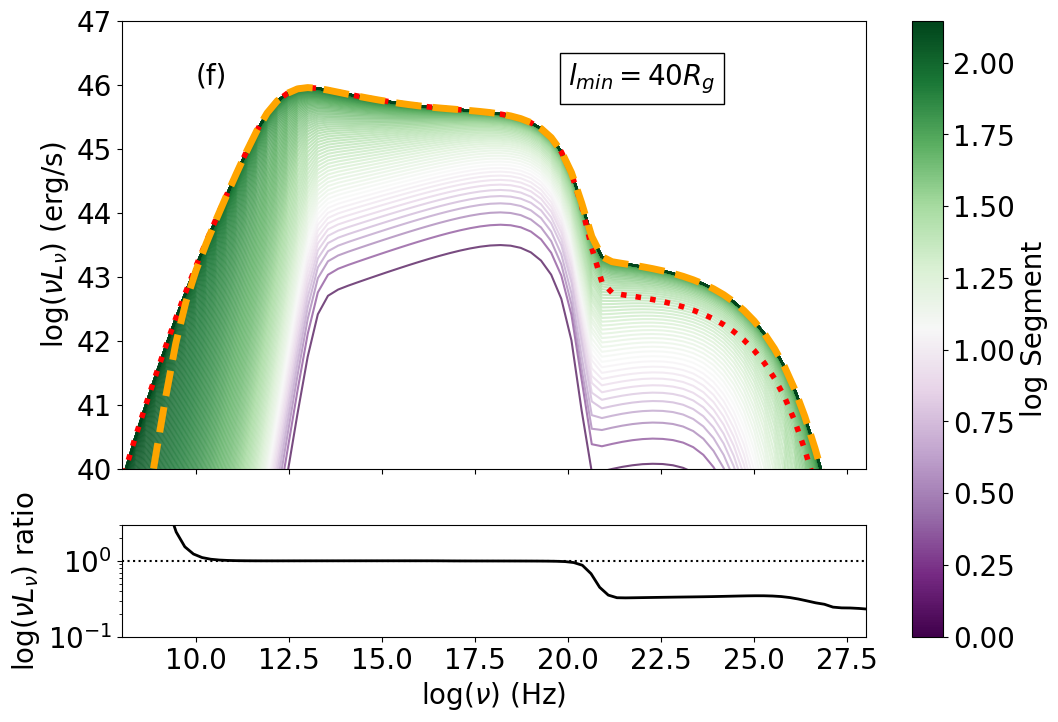}
\caption{Spectral energy distribution (SED) of a blazar jet with continuous magnetic dissipation for various model parameters. Starting from the baseline model in panel (a) of Fig.~\ref{multi-sed}, the panels here show models calculated for an additional range of altered parameters: a) electron escape, b) $\eta_{\rm acc}=100$, c) electron retention from up to 6 upstream segments, d) segment width reduced to a third of its baseline values, e) $L=5\cdot10^{47}$~erg s$^{-1}$, and f) $l_{\min}=40 R_g$. In each panel, spectra from segments along the jet are plotted with different colors. The dashed orange line indicates the spectrum arising from the distance where the dissipation profile peaks. The dotted red line shows jet spectrum calculated without taking into account photon spillage between segments. The ratio of these two spectra (in logarithmic scale) is displayed in the bottom panel of each plot.
}
\label{multi-sed2}
\end{figure*}

\end{document}